\documentstyle[aps,prl,preprint,tighten,eqsecnum]{revtex}
\draft
\newcommand{\bc}{\begin{center}}
\newcommand{\ec}{\end{center}}
\newcommand{\nin}{\noindent}
\newcommand{\be}{\begin{equation}}
\newcommand{\ee}{\end{equation}}
\newcommand{\ba}{\begin{array}}
\newcommand{\ea}{\end{array}}

\newcommand{\dif}{{\rm d}}
\newcommand{\tw}{\tau_{\scriptscriptstyle W}}

\newcommand{\snM}{s_{n}^{\scriptscriptstyle M}}
\newcommand{\Ns}{\scriptscriptstyle N}
\newcommand{\MT}{\tilde{M}}
\newcommand{\HT}{\tilde{H}}

\begin{document}
\title{\bf Quantum Mesoscopic Scattering:\\
Disordered Systems and Dyson Circular Ensembles}

\author{Rodolfo A. Jalabert$^{(1,2)}$ and Jean-Louis Pichard$^{(1)}$ \\
$ $ }

\address{$^{(1)}$CEA, Service de Physique de l'Etat Condens\'{e},
Centre d'\'Etudes de Saclay, \\
91191 Gif-sur-Yvette Cedex, France}

\address{$^{(2)}$Division de Physique Th\'eorique
\footnote{Unit\'e de
Recherche des Universit\'es Paris XI et Paris VI associ\'ee au CNRS},
Institut de Physique Nucl\'eaire, 91406 Orsay Cedex, France }

\date{7 September 1994}
\maketitle
\begin{abstract}
{We consider elastic reflection and transmission of electrons by
a disordered
system characterized by a $2N\!\times\!2N$ scattering matrix $S$.
Expressing $S$ in terms of the $N$ radial parameters
and of the four $N\!\times\!N$ unitary matrices
used for the standard transfer matrix parametrization, we calculate
their probability distributions
for the circular orthogonal (COE) and unitary (CUE) Dyson ensembles.
In this parametrization, we explicitely compare the COE--CUE
distributions with those suitable for quasi--$1d$ conductors and
insulators. Then, returning to the usual eigenvalue--eigenvector
parametrization of $S$, we study the distributions of
the scattering phase shifts. For a quasi--$1d$ metallic system,
microscopic simulations show that the
phase shift density and correlation functions are
close to those of the circular ensembles. When quasi--$1d$
longitudinal localization breaks $S$ into two uncorrelated
reflection matrices, the phase shift form factor $b(k)$
exhibits a crossover from a behavior characteristic of two uncoupled
COE--CUE (small $k$) to a single COE--CUE behavior (large $k$).
Outside quasi--one dimension, we find that the phase shift density
is no longer uniform and $S$ remains nonzero after disorder averaging.
We use perturbation theory to calculate the deviations to the
isotropic Dyson distributions. When the electron dynamics is no
longer zero dimensional in the transverse directions,
small-$k$ corrections to the COE--CUE behavior of $b(k)$ appear,
which are reminiscent of
the dimensionality dependent non universal regime of
energy level statistics. Using a known relation between the
scattering phase shifts and the system energy levels, we
analyse those corrections to the universal random matrix behavior
of $S$ which result from $d$--dimensional diffusion on short time
scales.}

\end{abstract}


\narrowtext

\newpage

\section {Introduction}
\label{sec:Intro}

The discovery of universal conductance fluctuations
\cite{RevWebb,AUCF,lee1} (UCF characterizing
the sensitivity of the conductance of small metallic samples to a
change of the Fermi energy, magnetic field or impurity configuration)
have generated a sustained interest in quantum mesoscopic physics
since the mid eighties. Mesoscopic systems
have a size of the order of the electron phase-coherence length $L_{\phi}$,
i.e. a scale intermediate between single atoms (microscopic) and bulk
solids (macroscopic). First, it is in terms of quantum interference
effects between different multiple scattering paths that UCF has been
understood \cite{AUCF,lee1}. The universality of the phenomenon
(reproducible fluctuations of order $e^{2}/h$, independently of the
mean conductance) quickly lead physicists \cite{AS,IEPL}
to understand UCF as a signature of a more general universality
resulting from the eigenvalue correlations of random matrices.

The standard random matrix ensembles were introduced in the context of
Nuclear Physics \cite{Wigner,Dy,Porter,Mehta}
and later found to also describe the statistical properties of
quantum systems whose classical analogs are chaotic \cite{BGS,RevBoh}.
Contrary to a microscopic approach where the system hamiltonian $H$,
scattering matrix $S$ or transfer matrix $M$ result from a more or less
arbitrary distribution of the substrate potential, random matrix theory
(RMT) assumes for $H$, $S$ or $M$ statistical ensembles resulting from an
hypothesis of maximum randomness, given the system symmetries
(time reversal symmetry, spin rotation symmetry, current conservation)
plus a few additionnal constraints.

A direct relationship between conductance fluctuations and RMT comes from
the Thouless expression of the conductance

\begin{equation}
G =  \frac{e^2}{\hbar} \ N(E_{c}) \ ,
\label{eq:Thoul}
\end{equation}

\nin given by the number of one-electron
\footnote {Throughout this work we will treat spinless electrons and
therefore we will not write spin-degeneracy factors.} levels
$N(E_{c})$ which lie within an energy band of width $E_{c}=D\hbar/L^2$
centered around the Fermi energy $E_f$.
The Thouless energy $E_{c}$ is the inverse characteristic time for an
electron to diffuse through the sample (of size $L$). The electron diffusion
coefficient is $D=v_{f}l/d$, where $v_{f}$ is the Fermi velocity, $l$ the
elastic mean free path, and $d$ the spatial dimension of the sample.
Fluctuations in the number of levels within $E_c$ are
therefore related to conductance fluctuations. The analysis of fluctuations
in the spectrum of non interacting electrons in metallic particles was
initiated by Gorkov and Eliashberg \cite{GoEl}. The relevance of RMT was
proved by Efetov \cite{Efe}
and complemented by Altshuler and Shklovski\u{\i} \cite{AS} who noticed non
RMT behavior for energy separations larger than $E_{c}$.
Using a diagramatic perturbation theory for the density-density
correlation function in the weak disorder limit $kl \gg 1$ ($k$ is
the Fermi wave vector), they showed that the correlation
between energy levels is correctly described by the universal RMT laws
for energy scales smaller than $E_{c}$, but are weaker (and dimensionality
dependent) beyond $E_c$.\footnote{This limit of the universal
RMT laws is specifically derived ignoring electron--electron interaction.}

 Pionieered by Imry \cite{IEPL}, an alternative approach relating UCF
and random matrix theories is based on the Landauer
formula, which gives the conductance in terms of the total
transmission coefficient of the sample (at the
Fermi energy), considered as
a single, complex elastic scatterer. For a two-probe measurement where the
disordered sample is attached between two perfect reservoirs with an
infinitesimal difference in their electrochemical potentials, the conductance
(measured in units of $e^2/h$) is

\begin{equation}
g = Tr[t t^{\dagger}] \ .
\label{eq:Land}
\end{equation}

\nin The transmission matrix $t$ can be expressed in terms of the transfer
matrix $M$, for which a standard random matrix theory (``global approach'')
has been developed\cite{MPS,St}. A set of $N$ real positive parameters
describing the radial part of the $2N\!\times\!2N$ transfer matrix $M$
(precise definitions are given in the next section) are the relevant
``eigenvalues" in this approach, and their probability distribution is

\begin{equation}
P(\{\lambda_a\})=\exp \left(-\beta {\cal H} (\{\lambda_a\})\right) \ ,
\end{equation}

\begin{equation}
{\cal H} (\{\lambda_a\})= -\sum_{a<b}^N \log
\left | \lambda_a - \lambda_b \right |+\sum_{a=1}^N V(\lambda_a) \ .
\label{globalapproach}
\end{equation}

\nin This is the usual RMT Coulomb gas analogy with logarithmic pairwise
interaction, a system dependent confining potential $V(\lambda)$, and an
inverse temperature $\beta=1,2,4$ depending on the system symmetries. This
distribution corresponds to the most random statistical ensemble for $M$
given the average density $\rho(\lambda)$ of radial parameters, which
controls the average conductance $\langle g \rangle$. In this maximum
entropy ensemble, $V(\lambda)$ and $\rho(\lambda)$ are related
by an integral relation in the large $N$--limit\cite{Dy2}:

\begin{equation}
V(\lambda)+ C = \int_0^{\infty} \dif \lambda' \rho(\lambda')
\log \left | \lambda - \lambda' \right | + { \beta -2 \over 2\beta}
\log \rho(\lambda) \ .
\label{Potential}
\end{equation}

\nin $C$ is a constant, and to leading order in $N$ this mean field equation
expresses the equilibrium of a charge at $\lambda$ resulting from its
interaction with the remaining charges and the confining potential.

 In this work we examine the statistical properties of another matrix related
to quantum transport in disordered systems: the scattering matrix $S$.

 First, the relation between $M$ and $S$ is straightforward, since
$M$ can be expressed in terms of the reflection and transmission
submatrices which define $S$. This allowes us to show that the
$\lambda$--statistics characterizing
the Dyson ensembles coincides with a ``global approach'' for $M$,
given a particular confining potential $V(\lambda)$. Using this
equivalence, the detailed proof of it being given in this work, the quantum
transport properties associated
with the circular ensembles have been obtained\cite{ropibe2}, including
the weak--localization effects and the quantum fluctuations of
various linear statistics of the $\lambda$--parameters (conductance,
shot--noise power, conductance of a normal--superconducting microbridge, etc).
This equivalence has been also derived by Baranger and
Mello \cite{meba}, and has been confirmed in their numerical simulations
of chaotic billiards.

 Second, since $S$ is unitary, its $2N$ eigenvalues $\{\exp{(i\theta_m)}\}$ are
just
given by the $2N$ phase shifts $\{\theta_m\}$. In the Dyson circular
ensembles, where all matrices $S$ with  a given symmetry are equally probable,
the phase shift statistics follow the same universal level correlations
than in the gaussian ensembles of corresponding symmetry. The applicability of
Dyson circular ensembles for chaotic scattering has been established
by the pioneering work of Bl\"umel and Smilansky \cite{BlSm}, as far as
the two-level form factor for the phase shifts is concerned.
Further evidence comes from numerical
studies of $2 \! \times \! 2$ matrices describing the scattering through
chaotic cavities \cite{DoroSm}.

 But a ballistic chaotic cavity,  with a conductance of order
$N/2$ and where the electronic motion is
essentially zero--dimensional after a (short) time of flight,
differs from a disordered conductor or insulator. The introduction of
bulk disorder in the dot reduces the conductance to smaller
values of order $Nl/L$ ($L >> l$ , metallic regime) and enhances
the time interval in which the electron motion depends on
system dimensionality.  The diffusive motion of the carriers
after a (short) characteristic time $\tau_e$ limits (rather than
improves) the validity of the standard RMT distributions. The
subject of this study consists precisely in understanding this apparent
paradox: the fact that the introduction in a cavity of bulk disorder
drives the
statistics of $S$ away from the circular ensembles. For this
purpose, we extensively study disordered
systems where non--interacting electrons are elastically scatterered
by  microscopic impurities contained in a rectangular dot of various
aspect ratios. We derive useful relations
between different parametrizations of $H$, $M$ and $S$, involving the
energy levels $\{E_i\}$, the radial parameters $\{\lambda_a\}$ and the
phase shifts $\{\theta_m\}$.

 The paper is organized as follows. After presenting the
basic definitions of our model and the relationship between
$S$ and $M$, we derive (Sec. \ref{sec:RelSM}) the metric
of $S$ in the polar decomposition, and therefore the probability
distribution of the radial parameters assuming a Dyson distribution for
the scattering matrix in the time-symmetric case (the case without
time-reversal symmetry is worked out in Appendix \ref{sec:CUE}). In
Sec. \ref{sec:disc} we use this parametrization to
discuss the differences and similarities between
the circular ensemble distributions and those suitable for
quasi-one dimensional disordered systems.
In Sec. \ref{sec:metal1} we start the statistical
analysis of the phase shifts of $S$ for long conductors with weak
disorder.
In Sec. \ref{sec:isol1} we study the scattering phase shifts in quasi-$1d$
insulators and show that their correlations are well described by those of
two uncoupled circular ensembles. In Sec. \ref{sec:Mean} we calculate by
diagrammatic perturbation theory the mean values of the $S$-matrix, which are
needed to understand the phase shift anisotropy found outside the quasi--$1d$
limit. In Sec. \ref{sec:Unfol}
we present numerical results for geometries other than quasi-one dimensional
strips (i.e. squares and thin slabs) and show that after the spectrum is
unfolded there is a rough agreement with the circular ensembles, though
noticable discrepancies are visible in the number variance. These deviations
are analyzed in Sec. \ref{sec:Devi}, using the known deviations of metallic
spectra
from the R.M.T. behaviors and exploiting a relationship between
scattering phase shifts and the energy levels (Appendix \ref{sec:Rela}).
Appendix
\ref{sec:Summa} gives a summary of the methods and results of the numerical
simulations, and in Appendix
\ref{sec:Semi}, we use a semiclassical approach to complement our
understanding of the results, including the meaning of the
Wigner time.

\section {Scattering, Transfer, Transport and ${\bf \lambda}$--parameters}
\label{sec:Scat}

Using only the system symmetries (current conservation, time reversal symmetry,
spin rotation symmetry) one can show that both $S$ and $M$ can be expressed
in terms of the $N$ radial parameters $\{\lambda_a\}$ of $M$ and 4
(2 in the presence of time reversal symmetry) $N\!\times\!N$ auxiliary unitary
matrices. In this parametrization, we shall show the similarities and the
differences
between the distributions implied by the circular ensembles and those
describing in the weak scattering limit long quasi--$1d$ disordered
conductors and insulators. To make concrete our explanations, we introduce in
what follows the essential elements of the particular microscopic model
on which we test the validity of Dyson circular ensembles.

We consider an infinite strip composed of two semi-infinite perfectly
conducting leads of width $L_{y}$ connected by a disordered part
of same width and of longitudinal length $L_{x}$ (Fig. 1). Assuming
non-interacting electrons and hard-wall boundary conditions for the
transverse part of the wavefunction, the scattering states in the leads
at the Fermi energy $E_{f}=\hbar^2 k^2/2m$ satisfy the condition
$k^2 = (n \pi/L_{y})^2 + k^{2}_{n}$, where $k$ is the
Fermi wavevector, $n \pi/L_{y}$ the quantized transverse wavevector and
$k_{n}$ the longitudinal momentum. The various transverse momenta labeled
by the index $n$ ($n=1,\ldots,N$) which satisfy this relationship with
$k_{n}^2 > 0$ define the $N$ propagating channels of the leads. Since
each channel can carry two waves travelling in opposite directions,
asymptotically far from the disordered region the wave function
can be specified by a $2N$-component vector on the two sides of the
disordered part. In each lead the first $N$ components are the amplitudes
of the waves propagating to the right and the remaining $N$ components are
the amplitudes of the waves travelling to the left.

\begin{mathletters}
\label{allwfs}
\begin{equation}
\Psi_{I}(x,y) = \sum_{n=1}^{N} \frac{1}{k_{n}^{1/2}} \left(A_{n} e^{i k_{n} x}
+ B_{n} e^{-i k_{n} x}\right)\phi_{n}(y) \ ,
\label{wfa}
\end{equation}
\begin{equation}
\Psi_{II}(x,y) = \sum_{n=1}^{N} \frac{1}{k_{n}^{1/2}} \left(C_{n}
e^{i k_{n}(x-L_{x})}
+ D_{n} e^{-i k_{n}(x-L_{x})}\right)\phi_{n}(y) \ .
\label{wfb}
\end{equation}
\end{mathletters}

\nin The transverse wavefunctions are
$\phi_{n}(y)=\sqrt{2/L_{y}} \ \sin{(\pi n y/L_{y})}$. The normalization is
chosen in order to have a unit incoming flux in each channel.
The scattering matrix
$S$ relates the incoming flux to the outgoing flux

\begin{equation}
\left( \begin{array}{l}
B \\
C
\end{array} \right)
= S \
\left( \begin{array}{l}
A \\
D
\end{array} \right) \ .
\label{eq:Sdef}
\end{equation}

\nin $S$ is a $2N \! \times \! 2N$ matrix of the form

\begin{equation}
S = \left( \begin{array}{lr}
r	& \hspace{0.5cm} t'	\\
t	& \hspace{0.5cm} r'
\end{array} \right) \ .
\label{eq:Smat}
\end{equation}

\nin The reflection (transmission) matrix $r$ ($t$) is an $N \times N$
matrix whose elements $r_{ba}$ ($t_{ba}$) denote the refected (transmitted)
amplitude in channel $b$ when there is a unit flux incident from the left
in channel $a$. The amplitudes $r'$ and $t'$ have similar meannings, except
that the incident flux comes from the right.
\footnote {Throughout this work we shall frequently represent
$2N \! \times \! 2N$ matrices in terms of their $N \! \times \! N$ blocks. We
reserve capital letters for $2N \! \times \! 2N$ matrices, while
calligraphic and low-case letters are used for $N \! \times \! N$ matrices.}
Current conservation
implies that $S$ is unitary.
Note that, with the convention we have taken, $S$ for a
perfect (non-disordered) sample at zero magnetic field is not the
identity matrix but is characterized by transmission submatrices which
contain pure phases $t_{ba}=t'^*_{ba}= \delta_{ab}\exp{(ik_{b}L_{x})}$.

The $2N \! \times \! 2N$ transfer matrix relates the flux amplitudes on
the left-hand side of the disorder part with those on the right:

\begin{equation}
\left( \begin{array}{l}
C \\
D
\end{array} \right)
= M \
\left( \begin{array}{l}
A \\
B
\end{array} \right) \ .
\label{eq:Mdef}
\end{equation}

\nin Just as for $S$, we can write $M$ in terms of four $N \! \times \! N$
blocks

\begin{equation}
M = \left( \begin{array}{lr}
m_{1}	& \hspace{0.5cm} m_{2}	\\
m_{3}	& \hspace{0.5cm} m_{4}
\end{array} \right) \ .
\label{eq:Mmat}
\end{equation}

The reflection and transmission matrices of $S$ can be
expressed in terms of the block matrices $m_{i}$ of $M$.
Introducing the polar representation\cite{MPK,mellopichard} of $M$,
we have:

\begin{mathletters}
\label{allrels}
\begin{equation}
r = - m_{4}^{-1}m_{3} = - u^{(3)} {\cal R} u^{(1)} \ ,
\label{rela}
\end{equation}
\begin{equation}
t = (m_{1}^{\dagger})^{-1} = u^{(4)} {\cal T} u^{(1)} \ ,
\label{relb}
\end{equation}
\begin{equation}
r' = m_{2} m_{4}^{-1} = u^{(4)} {\cal R} u^{(2)} \ ,
\label{relc}
\end{equation}
\begin{equation}
t' = m_{4}^{-1} = u^{(3)} {\cal T} u^{(2)} \ ,
\label{reld}
\end{equation}
\end{mathletters}

\nin where $u^{(l)}$ ($l=1,\ldots,4)$ are arbitrary $N \! \times \! N$ unitary
matrices. ${\cal R}$ and ${\cal T}$ are real diagonal $N \! \times \! N$
matrices whose non-zero elements (labeled by only one index) are the
square roots of the
reflection and transmission eigenvalues which can be expressed as a
function of the real positive diagonal elements $\lambda_{a}$
($a=1,\ldots,N$) of the $N \! \times \! N$ diagonal matrix $\lambda$
characterizing the radial part of $M$:

\begin{mathletters}
\label{allrtcals}
\begin{equation}
{\cal R}_{a} = \left(\frac{\lambda_{a}}{1+\lambda_{a}}\right)^{1/2} \ ,
\label{rtcala}
\end{equation}
\begin{equation}
{\cal T}_{a} = \left(\frac{1}{1+\lambda_{a}}\right)^{1/2} \ .
\label{rtcalb}
\end{equation}
\end{mathletters}

\nin In this $\lambda$--parametrization, $M$ and $S$ can then
be written as

\begin{equation}
M = \left( \begin{array}{cc}
u^{(4)}	& \ 0	\\
0    	& \ u^{(2) \dagger}
\end{array} \right)
 \left( \begin{array}{cc}
(I+\lambda)^{1/2} 	& \hspace{0.5cm} \lambda^{1/2} \\
\lambda^{1/2} 		& \hspace{0.5cm} (I+\lambda)^{1/2}
\end{array} \right)
 \left( \begin{array}{cc}
u^{(1)}	& \ 0	\\
0	& \ u^{(3) \dagger}
\end{array} \right) \ ,
\label{eq:Mpol}
\end{equation}

\begin{equation}
S = \left( \begin{array}{cc}
u^{(3)}		& \ 0	\\
0		& \ u^{(4)}
\end{array} \right)
 \left( \begin{array}{cc}
-{\cal R}	& \hspace{0.5cm} {\cal T}	\\
{\cal T}	& \hspace{0.5cm} {\cal R}
\end{array} \right)
 \left( \begin{array}{cc}
u^{(1)}	& \ 0	\\
0	& \ u^{(2)}
\end{array} \right) \ .
\label{eq:Spol}
\end{equation}

Since $t t^{\dagger} = u_{1} {\cal T}^2 u_{1}^{\dagger}$, the
dimensionless conductance can be expressed as

\begin{equation}
g = \sum_{a=1}^{N} \ \frac{1}{1+\lambda_{a}} \hspace{0.5cm} .
\label{eq:geigen}
\end{equation}

\nin The conductance is therefore a linear statistics of the radial
parameters $\{\lambda_{a}\}$ of $M$. Note that such a simple
relationship does not exists between $g$ and the scattering
phase--shifts $\{\theta_a\}$, since the relation between the eigenvalues
of $S$ and those of $t t^{\dagger}$ depends on the eigenvectors of
$S$ and on the $u$--matrices.

In the absence of a magnetic field there is time reversal symmetry,
the $S$-matrix is symmetric ($S = S^{\rm T})$ and the polar decomposition
has only two arbitrary unitary matrices since

\begin{mathletters}
\label{allB0s}
\begin{equation}
u^{(3)} = u^{(1) \rm T} \ ,
\label{B0a}
\end{equation}
\begin{equation}
u^{(4)} = u^{(2) \rm T} \ .
\label{B0b}
\end{equation}
\end{mathletters}

\nin In this case the number of independent parameters is to $2N^{2}+N$. We
have $N^{2}$ parameters for each of the two $N \! \times \! N$ unitary
matrices and $N$ for the diagonal matrix $\lambda$. Without time-reversal
symmetry (unitary case with spin degeneracy) the number of
independent parameters of $S$ (and $M$) is $4N^{2}$ (the $N$ extra parameters
of the polar decompositions (\ref{eq:Spol}) and (\ref{eq:Mpol}) are due to
the fact that they are not unique \cite{St}).
In the symplectic case occuring when there is a strong spin--orbit
scattering in the disordered part and no applied magnetic field,
the spin degenracy is removed and
each matrix element becomes a $2 \times 2$ quaternion matrix,
which doubles the size of $M$ and $S$, but
$u^{(3)}$ and $u^{(4)}$ are also given\cite{mellopichard} in terms of
$u^{(1)}$ and $u^{(3)}$ and the $\lambda$ have a twofold degeneracy
(Kramers degeneracy).

\section {Invariant measure of $S$ in the polar decomposition}
\label{sec:RelSM}

In this section we calculate the invariant measure $\mu(dS)$ of $S$ in
terms of the radial parameters $\{\lambda_{a}\}$ and the matrices
$u^{(l)}$. We present here the time-symmetric case ($\beta=1$), while
the unitary case ($\beta=2$) relevant when a magnetic field is applied
is considered in Appendix \ref{sec:CUE}. Our calculations have recently been
extended by K.~Frahm \cite{KF} for the symplectic case ($\beta=4$).
In the orthogonal case $S$ is unitary symmetric and can be decomposed as:

\begin{equation}
S = W^{\rm T} \Sigma \ W = U^{\rm T} \ \Gamma \ U = Y^{\rm T} \ Y \ .
\label{eq:Sdecomp}
\end{equation}

\nin The first equality simply means the diagonalization of $S$ and
introduces the $2N$ phase shifts of $S$ through the diagonal elements
$\exp{(i\theta_{m})}$ of $\Sigma$ and a $2N \times 2N$ orthogonal matrix
$W$ containing  the eigenvectors of $S$. The second equality
results of the polar representation of $S$ where the real matrix $\Gamma$
and the block-diagonal unitary matrix $U$ are given by Eq.~(\ref{eq:Spol})
with the conditions (\ref{allB0s}). The last decomposition holds for any
unitary symmetric matrix and introduces a $2N \times 2N$ unitary matrix
$Y$ which is not unique, but specified up to an orthogonal transformation.
Following Dyson\cite{Dy}, the measure of a neighbourhood $dS$ of $S$ is
given in terms of the infinitesimal variations $d\MT_{ij}$ of a the matrix
elements of a real symmetric matrix $d{\MT}$ defined by:

\begin{equation}
dS = Y^{\rm T} \ (id\MT) \ Y \ ,
\label{eq:dS}
\end{equation}

\begin{equation}
\mu(dS) = \prod_{i \leq j}^{2N} d\MT_{ij} \ .
\label{eq:mudS}
\end{equation}

\nin This definition is {\it independent} of the particular choice of the
unitary matrix $Y$ and we use this freedom of choice to take a convenient
$Y$ for expressing $d\MT$ in the $\lambda$--parametrization. To this end,
we note that $\Gamma$ is real symmetric and unitary, with eigenvalues
$\pm 1$ and diagonalizable by an orthogonal transformation $O$:

\begin{equation}
\Gamma = O^{\rm T} \ D \ O \ ,
\label{eq:Gamma}
\end{equation}

\begin{equation}
D = \left( \begin{array}{cc}
I	& \hspace{0.5cm} 0 \\
0	& \hspace{0.5cm} -I
\end{array} \right) \hspace{0.5cm} , \hspace{2.5cm}
O = \left( \begin{array}{cc}
{\cal P}	& \hspace{0.5cm} {\cal Q} \\
{\cal Q}	& \hspace{0.5cm} -{\cal P}
\end{array} \right) \ .
\label{eq:Omat}
\end{equation}

\nin The $N \! \times \! N$ blocks of $O$ are diagonal matrices given by

\begin{equation}
{\cal P}_{a} = \frac {1}{\sqrt{2}} \ \sqrt{1-{\cal R}_{a}} \hspace{0.5cm} ,
\hspace{2.0cm} {\cal Q}_{a} = \frac {1}{\sqrt{2}} \ \sqrt{1+{\cal R}_{a}} \ .
\label{obla}
\end{equation}

\nin Writing the diagonal matrix $D$ as $F^{2}$, with

\begin{equation}
F = \left( \begin{array}{lr}
I & \hspace{0.5cm} 0 \\
0	& \hspace{0.5cm} iI
\end{array} \right) \ ,
\label{eq:Fmat}
\end{equation}

\nin one can write $S$ as $Y^{\rm T} Y$, with

\begin{equation}
Y = F \ O \ U = \left( \begin{array}{lr}
{\cal P}u^{(1)} & \hspace{0.5cm} {\cal Q}u^{(2)} \\
i{\cal Q}u^{(1)}	& \hspace{0.5cm} -i{\cal P}u^{(2)}
\end{array} \right) \ .
\label{eq:Umat}
\end{equation}

\nin Since $Y$ is unitary, its infinitesimal variations can be
expressed as

\begin{equation}
dY = \delta Y \ Y \ ,
\label{eq:dU}
\end{equation}

\nin where the matrix $\delta Y$ is antihermitic. Analogously, for the bock
components of $U$ we can write

\begin{equation}
du^{(l)} = \delta u^{(l)} \ u^{(l)} \ , \hspace{2cm} \delta u^{(l)} =
da^{(l)}+i \ ds^{(l)} \ ,
\hspace{2cm} l=1,2 .
\label{eq:dvl}
\end{equation}

\nin $d a^{(l)}$ ($d s^{(l)}$) are real antisymmetric (symmetric)
$N \! \times \! N$ matrices. The Haar measure $\mu (d u^{(l)})$ for
the unitary matrices $u^{(l)}$ satisfies $\mu (du^{(l)}) =
\prod_{a} ds^{(l)}_{aa} \prod_{a<b}da^{(l)}_{ab} ds^{(l)}_{ab}$.
Therefore the infinitesimal variations of $Y$ and $S$ are given by

\begin{eqnarray}
\delta Y = \left( \begin{array}{cc}
0 & \hspace{0.5cm} i \ (d{\cal Q} \ {\cal P} - d{\cal P} \ {\cal Q}) \\
i \ (d{\cal Q} \ {\cal P} - d{\cal P} \ {\cal Q}) & 0
\end{array} \right) +
\nonumber \\
\nonumber \\
+ \left( \begin{array}{cc}
{\cal P} \ \delta u^{(1)} \ {\cal P}+{\cal Q} \ \delta u^{(2)} \ {\cal Q} &
\hspace{0.5cm}
i \ (- {\cal P} \ \delta u^{(1)} \ {\cal Q} + {\cal Q} \ \delta u^{(2)}
\ {\cal P}) \\
i \ ({\cal Q} \ \delta u^{(1)} \ {\cal P}-{\cal P} \ \delta u^{(2)} \ {\cal Q})
& \hspace{0.5cm} {\cal Q} \ \delta u^{(1)} \ {\cal Q}+{\cal P} \ \delta u^{(2)}
\ {\cal P}
\end{array} \right)  \ ,
\label{eq:dAmat}
\end{eqnarray}

\begin{equation}
dS =  Y^{\rm T}(\delta Y - \delta Y^{*})Y = Y^{\rm T}(id\MT)Y \ .
\label{eq:dUftm}
\end{equation}

\nin This give us the real symmetric matrix $d\MT$ in terms of the radial
parameters
$\{\lambda_a\}$, the unitary matrices $u^{(l)}$ and their infinitesimal
variations $\{d\lambda_a\}$ and $\delta u^{(l)}$. We just need to calculate
a Jacobean which can be decomposed as the product of three determinants,

\begin{mathletters}
\label{allbtps}
\begin{equation}
\prod_{a=1}^{N} d\MT_{a,a+\Ns} = \prod_{a=1}^{N}
\frac{1}{2 \sqrt{\lambda_{a}} (1+\lambda_{a})} \ d \lambda_{a} \ ,
\label{btpa}
\end{equation}
\begin{equation}
\prod_{a<b}^{N} d\MT_{a,b+\Ns} \ d\MT_{b,a+\Ns} = \prod_{a<b}^{N} \ 2 \
\left(\sqrt{\frac{\lambda_{a}}{\lambda_{a}+1}} -
\sqrt{\frac{\lambda_{b}}{\lambda_{b}+1}} \right)
da^{(1)}_{ab} da^{(2)}_{ab} \ ,
\label{btpb}
\end{equation}
\begin{equation}
\prod_{a \leq b}^{N} d\MT_{a,b} \ d\MT_{a+\Ns,b+\Ns} =
\prod_{a=1}^{N} 4 \ \sqrt{\frac{\lambda_{a}}{\lambda_{a}+1}}
ds^{(1)}_{aa} ds^{(2)}_{aa} \
\prod_{a<b}^{N} \ 2 \
\left(\sqrt{\frac{\lambda_{a}}{\lambda_{a}+1}} +
\sqrt{\frac{\lambda_{b}}{\lambda_{b}+1}} \right)
ds^{(1)}_{ab} ds^{(2)}_{ab} \ ,
\label{btpc}
\end{equation}
\end{mathletters}

\nin which eventually gives for the invariant measure of the symmetric unitary
matrix $S$

\begin{equation}
\mu(dS) = \prod_{a=1}^N \frac{1}{(1+ \lambda_a)^{3/2}} \
\prod_{a,b}^N {\left| {1 \over 1+ \lambda_a} -
{ 1 \over 1+\lambda_b} \right |} \
\mu(d\lambda) \ \prod_{l=1}^2 \mu(du^{(l)})
\label{eq:mudSf}
\end{equation}

\nin in terms of the measure $\mu (d\lambda) =\prod_{a=1}^{N} d\lambda_{a}$ of
the matrix $\lambda$ and of the Haar measures $\mu(du^{(l)})$ of the
matrices $u^{(l)}$.

\section {Dyson circular ensembles and Quasi--$1d$ disordered systems}
\label{sec:disc}

For the Dyson circular ensembles the number of $S$--matrices in a volume
element $dS$ of measure $\mu(dS)$ around a given $S$ is just proportional to
$\mu(dS)$ :

\be
P(dS)={1 \over V} \ \mu(dS) \ ,
\ee

\nin $V$ is a normalization constant. Using the $\lambda$--parametrization,
one obtains that the matrices $u^{(l)}$ are independent from each
other (except by symmetry relations) and distributed according the invariant
Haar measure on the unitary group, while the $N$ parameters $\lambda_a$
are statistically independent from the $u$--matrices and have a joint
probability distribution which can be expressed in the usual Coulomb gas
analogy as a Gibbs function

\begin{equation}
P(\{\lambda_{a}\}) = \exp{(- \beta {\cal H}(\{\lambda_{a}\}))} \ .
\label{eq:jacobian}
\end{equation}

\nin The symmetry parameter $\beta$ plays the role of an inverse
temperature and the effective hamiltonian ${\cal H}$ is characterized by a
logarithmic pairwise interaction and a one-body potential:

\begin{equation}
{\cal H}(\{\lambda_{a}\}) = \sum_{a<b}^{N} f(\lambda_{a},\lambda_{b}) +
\sum_{a=1}^{N} V(\lambda_{a}) \ ,
\label{eq:hamiltonian}
\end{equation}

\begin{equation}
f(\lambda_{a},\lambda_{b}) = - \log{\left|\lambda_{a} - \lambda_{b} \right|}
\ ,
\label{eq:interaction}
\end{equation}

\begin{equation}
V(\lambda) = \left( N + {\beta - 2 \over 2\beta}
 \right) \ \log{(1+\lambda)} \ .
\label{eq:potential}
\end{equation}

 This Coulomb gas analogy characterizes the orthogonal, unitary and
symplectic ensembles which differ not only by the
value of the ``temperature'' $\beta^{-1}$, but also by the presence
of a small $\beta$--dependent correction to the leading behavior
of $V(\lambda)$ in a large $N$--expansion. It is remarkable that
the pairwise interaction for the $\lambda$--parameter is the same
than in the global maximun entropy approach
to the transfer matrix \cite{St}, while $V(\lambda)$ differs in two
important aspects: it is essentially
proportional to $\log{\lambda}$ instead of $\log^{2}{\lambda}$ \cite{Sle}
for large values of $\lambda$, and the prefactor is just the number of
modes $N$ instead of the classical conductance $Nl/L_{x}$. These
differences are not surprising since forward and backward scattering are
essentially put on the same footing in the the circular ensembles, leading
to a total transmission intensity of the order of the total reflection
intensity, $T \approx  R \approx N/2$ (up to weak--localization
corrections) \cite{ropibe2,meba}. For a disordered conductor or insulator,
the refection $R$ is
much larger than the transmission $T$. Clearly, bulk diffusion characterized
by an elastic mean-free-path $l$ cannnot be described by the circular
ensembles, which are appropriate for systems where an injected
carrier is subjected to a chaotic dynamics
before finding [with equal chance] one of the two injection leads.
However, as shown by Beenakker\cite{Been2}, in the large $N$--limit, the
$\lambda$ density--density correlation function (and therefore the variance
of any linear statistics like the conductance) depends only on the pairwise
interaction, and not on the particular form for $V(\lambda)$.
Consequently, both the circular ensembles and the global maximum
entropy approach to the transfer matrix yield identical UCF \cite{PMPC} values
$2 / (16 \beta)$, slightly different from the perturbative microscopic result
\cite{AUCF,lee1} for quasi--$1d$ disordered conductors $2 / (15 \beta)$.

In the polar representation of $S$, one can precisely see the difference
between the circular ensembles and those appropriate for quasi--one
dimensional disordered systems. For this, we just need to recall what we
know from another statistical approach introduced for arbitrary $N$ by
Dorokhov\cite{Dorok} from microscopic considerations and by Mello
{\it et al} \cite{Mello,MPK}
from a maximum entropy assumption for the infinitesimal transfer
matrix of the building block of a quasi--$1d$ series. These works are
based on an isotropy hypothesis: it is assumed that
the $u$--matrices are distributed with the Haar measure on the unitary
group and statistically independent from the radial part of $M$
(see Eq. \ref{eq:Mpol}). This limits their conclusions
to quasi--one dimension and yields a Fokker-Planck equation for
$P(\{\lambda_a\})$ which implies the same UCF and weak-localization
corrections for quasi--$1d$ conductors than those given by diagrammatic
calculations.
The evolution of $P(\{\lambda_a\})$ with the length
$L_x$ of the disordered part is given by an heat equation where the Laplacian
becomes the radial part of the Laplace-Beltrami operator on a space of
negative curvature. Using Sutherland's transformation, Beenakker and Rejaei
\cite{Been} have mapped this diffusion equation into a Schr\"odinger equation
(with imaginary time) of
a quantum set of point like particles free to move on a half line (the
positive part of the real axis) within a certain potential. For
arbitrary values of $\beta$, these particles have a
pairwise interaction, attractive for $\beta=1$ and repulsive for $\beta=4$,
making difficult to find the solution.
Fortunately, this interaction vanishes for $\beta=2$, and the solution of
the diffusion equation is reduced to an exactly solvable quantum $N$--body
free fermion problem. This gives for the unitary case a pairwise interaction

\begin{equation}
f(\lambda_a-\lambda_b)= - \ \frac{1}{2} \ \ln \left|\lambda_a-\lambda_b\right|
\ - \ \frac{1}{2} \ \ln \left|{\rm arcsinh}^2 \left(\sqrt{\lambda_a}\right) -
{\rm arcsinh}^2 \left(\sqrt{\lambda_b}\right)\right| \ ,
\end{equation}

\nin which reduces to the usual logarithmic interaction assumed by the
global approach for $|\lambda_a - \lambda_b| \ll 1 $, but which is halved if
$|\lambda_a - \lambda_b| \gg 1$ in the quasi--$1d$ diffusive
or {\it localized} limit. This discrepancy is responsible for the slightly
different UCF values characterizing ballistic quantum dots with chaotic
dynamics and quasi--$1d$ disordered conductors.

In the localized regime, the global and local approaches give
identical symmetry dependence of the localization lengths, though the (log)
conductance fluctuations differs by a factor 2 in the quasi--$1d$ localized
limit\cite{Pichard}. This later point again is consistent with the halving of
the pairwise interaction $f(\lambda_a - \lambda_b)$ for large eigenvalue
separations given by the local approach. For metals and
insulators far from a quasi--one dimensional shape, a more dramatic
shrinkage of the validity of the universal RMT--correlations has been
observed \cite{Sle}. This means that transverse diffusion
(or even more transverse localization\cite{AviPicMut}) yields a more
significant reduction of the RMT pairwise interaction than the one obtained
in quasi-one dimension by Beenakker and Rejaei. To address this
problem, we return to the study of the more familiar scattering
phase shifts of $S$.

\section {Scattering phase shifts in quasi--$1d$ metals and a single
circular ensemble.}
\label{sec:metal1}

As discussed in the previous section, when applied to
quasi--$1d$ conductors, the circular ensembles give the
right statistics for the $u$--matrices, and locally the correct
interaction $f(\lambda_a, \lambda_b)$, but certainly not the appropriate
confining potential $V(\lambda)$. Using now the eigenvalue--eigenvector
parametrization of $S$, we study the phase shift
distribution in the case of weak disorder and  quasi--$1d$ samples. We first
introduce the basic notation, then write the expected universal correlation
for Dyson ensembles, which we compare to our numerical results.

Since the $S$-matrix is unitary, its $2N$ eigenvalues are given by $2N$ phase
shifts $\theta_m$. Following Bl\"umel and
Smilansky \cite{BlSm} we write the phase shift density as

\begin{equation}
\rho(\theta) = \sum_{m=1}^{M} \langle \delta (\theta -\theta_{m})
\rangle =
\frac{1}{2\pi} \sum_{n=-\infty}^{\infty} \exp{(- i n \theta)}
\langle Tr S^n\rangle \ ,
\label{eq:dens}
\end{equation}

\nin where the angular brackets indicate average over the ensemble of
disordered samples and $M=2N$ is the dimension of $S$.
The two-point correlation function $R_{2}$ is defined by

\begin{equation}
R_{2}(\theta_{1},\theta_{2}) = \sum_{m\neq m'}^{M} \langle
\delta (\theta_{1} -\theta_{m})
\delta (\theta_{2} -\theta_{m'})
\rangle \ .
\label{eq:R2def}
\end{equation}

\nin When the phase shift distribution is uniform ($\rho(\theta)=M/2\pi$) the
two-point correlation function depends only on the difference
$\eta = \theta_{2} - \theta_{1}$,

\begin{equation}
R_{2}(\eta) =
\frac{M}{(2\pi)^2} \sum_{n=-\infty}^{\infty}
\left(\frac{1}{M}\langle |TrS^n|^2\rangle-1 \right)
\exp{(in\eta)} \ .
\label{eq:R2}
\end{equation}

The two-level cluster function is defined, for the reduced variable
$r = \eta M/2\pi$, in the limit where the number of
phases $M$ goes to infinity, as:

\begin{equation}
Y_{2}(r) = \lim_{M \rightarrow \infty} \hat{Y}_{2}^{M}(r) \ ,
\label{eq:Y2}
\end{equation}

\begin{equation}
\hat{Y}_{2}^{M}(r) = \left(\frac{2 \pi}{M}\right)^2 \left( \left(\frac{M}
{2 \pi}\right)^2 - R_{2} \left(\frac{2 \pi r}{M} \right) \right) \ .
\label{eq:Y2H}
\end{equation}

\nin Using the expression (\ref{eq:R2}) we have that

\begin{equation}
\hat{Y}_{2}^{M}(r) = \frac{1}{M} \left(1 - 2 \sum_{n=1}^{\infty} \snM
\cos{ \left(\frac{2 \pi n r}{M}\right)}\right) \ ,
\label{eq:Y2F}
\end{equation}

\nin where Fourier components $\snM$ are given by

\begin{equation}
\snM = \frac{1}{M}
\langle |TrS^n|^2 \rangle - 1 \ .
\label{sn}
\end{equation}

The argument $r$ of the cluster function goes from $- \infty $ to $+ \infty $,
and the Fourier transform of $Y_{2}$, the two-level form factor (TLFF), is
given by

\begin{equation}
b(k) = \int_{-\infty}^{\infty} dr \ Y_{2}(r) \exp{(2 \pi i k r)} \ .
\label{eq:bdk}
\end{equation}

\nin Comparing the Fourier transform of $Y_{2}$ and the Fourier
coefficients of $\hat{Y}_{2}^{M}$ in the large $M$ limit, we can identify

\begin{equation}
\snM \approx - \ b(n/M) \hspace{0.5in} , \hspace{0.5in}   M \gg 1 \ .
\label{eq:equiv}
\end{equation}

For matrices $S$ belonging to Dyson ensembles, the distribution of phase
shifts is given by the Coulomb gas analogy:

\be
P(\{\theta_a\})=\frac{1}{Z} \exp{\left(-\beta {\cal H}\{\theta_a\}\right)} \ ,
\ee

\nin where the effective hamiltonian is
\be
{\cal H} (\{\theta_a\})=-\sum_{a<b}^{M}
\ \log \left|e^{i\theta_a}- e^{i\theta_b}\right| \ .
\ee

The cluster functions of the circular ensembles have universal
forms which depends only on the system symmetries. For the
unitary and orthogonal ensembles $Y_{2}$ is an even function of its argument
and has the form \cite{Mehta}
\footnote{We follow the standard notation: ${\rm Si}(x) = \int_{0}^{x}
\frac{\sin{y}}{y} \ dy$; ${\rm Ci}(x) = {\cal C} + \ln{x} + \int_{0}^{x}
\frac{\cos{y}-1}{y} \ dy$; $\epsilon(x) = - 1/2, 0, 1/2$, for $x<0$, $x=0$,
$x>0$ respectively; ${\cal C} = 0.5772\ldots$ is the Euler constant.}

\begin{mathletters}
\label{allequations}
\begin{equation}
Y_{2}^{UE}(r) = \left(\frac{\sin{\pi r}}{\pi r}\right)^2  \ ,
\label{equationa}
\end{equation}
\begin{equation}
Y_{2}^{OE}(r) = \left(\frac{\sin{\pi r}}{\pi r}\right)^2  -
\left({\rm Si}(\pi r) -
\pi \varepsilon(r)\right) \left(\frac{\cos{\pi r}}{\pi r} -
\frac{\sin{\pi r}}{(\pi r)^2} \right) \ .
\label{equationb}
\end{equation}
\end{mathletters}

The corresponding form factors are also even functions of their
argument, and have the universal forms:

\begin{mathletters}
\label{allbes}
\begin{equation}
b_{UE}(k) = \left\{ \begin{array}{ll}
1 - k  & \hspace{1cm} \mbox{if $k \leq 1$} \\
0 & \hspace{1cm} \mbox{if $k \geq 1$}
\end{array} \right. \ ,
\label{bea}
\end{equation}
\begin{equation}
b_{OE}(k) = \left\{ \begin{array}{ll}
1 - 2 k + k \ \ln{(1+2 k)} & \hspace{1cm} \mbox{if $k \leq 1$} \\
1 - k \ \ln{\left(\frac{2 k + 1}{2 k -1}\right)} & \hspace{1cm}
\mbox{if $k \geq 1$}
\end{array} \right. \ .
\label{beb}
\end{equation}
\end{mathletters}

We check for a quasi--$1d$ metal (Fig. 2) the agreement between
the numerically generated Fourier components $\snM$ and the
universal two-level form factors $b(k)$ of Eq.~(\ref{allbes}), assuming
Eq.~(\ref{eq:equiv}). The $S$--matrix of disordered strips described by
a tight-binding Anderson model of $34 \times 136$ sites (details given
in Appendix \ref{sec:Summa}) are numerically evaluated. Averaging involves
5000 different impurity configurations. The Fermi energy in units
of the constant off-diagonal hopping term is $E = - 2.5$, the number of
propagating modes is $N=14$ and therefore the dimension of $S$ is $M=28$.
A relatively low wave-vector ($k = 1.32$ in Anderson units) is taken in order
to avoid lattice effects,
since we will be interested in the comparison between our simulations
and analytical approaches (diagrammatic perturbation theory and
semiclassical approximation) assuming a continuum limit. One finds a
rather good agreement for the time-reversal symmetric case (run R1,
no magnetic field, COE-like) and for the non time-reversal symmetric case
(run F1, with magnetic field, CUE-like). The distribution of the phase shifts
(inset) is quite uniform with and without magnetic field. The disorder in
the samples is very weak ($W=1$ in units of the hopping term) giving
an elastic mean-free-path $l=0.7 L_{y}$, and an average conductance
$\langle g \rangle = 4.1$.

 We conclude that the phase shift density and correlations for a quasi--$1d$
{\it conductor} are well approximated by the corresponding COE--CUE density
and correlations. Since it is clear in the
$\lambda$--parametrization that a quasi--$1d$ conductor cannot be seen as
a member of a COE--CUE ensemble, we suspect that this numerical result
merely indicates a good approximation, and that the main non COE--CUE
behavior of $S$ for a quasi--$1d$ conductor must occur in the distribution of
its eigenvectors. As we will see in the following sections, the good
agreement with the universal correlations gets poorer as we increase the
disorder, enter into the quasi-$1d$ localized regime, or go outside the
quasi-$1d$ geometry.

\section {Scattering phase shifts in quasi--$1d$ insulators and two
uncoupled circular ensembles.}
\label{sec:isol1}

The approximate COE--CUE behavior which we found in the previous section
cannot remain in the presence of quasi--$1d$ localization for obvious reasons:
$g \ll 1$ and a typical
matrix element of the reflection matrices $r$ and $r'$
is much larger than those of $t$ and $t'$. The matrix $S$ can then be thought
as two diagonal blocks, $r$ and $r'$, weakly coupled by $t$ and $t'$.
Using the polar decomposition, Eq. (\ref{eq:Spol}), and the fact that the
radial parameters $\{\lambda_{a}\}$ are exponentially large in the localized
regime, one can write

\begin{mathletters}
\label{alllocs}
\begin{equation}
r = - u^{(3)}u^{(1)} + {\cal O}(\lambda^{-1})  \ ,
\label{loca}
\end{equation}
\begin{equation}
r' = u^{(4)}u^{(2)} + {\cal O}(\lambda^{-1}) \ .
\label{locb}
\end{equation}
\end{mathletters}

\nin A strong quasi--$1d$ localization implies that $S$ reduces to
two uncoupled $N \! \times \! N$ unitary (symmetric if the presence of
time reversal symmetry) matrices and isotropy means that each of them is
invariant under orthogonal (unitary) transformation. The phase shifts
associated to $r$ and $r'$ will then be described
separately by two uncoupled COE--CUE ensembles.
For a weaker quasi--$1d$ localization, we have a cross--over behavior
between a set of $2N$ phase shifts with approximately COE--CUE correlations
to two uncoupled sets of $N$ exactly COE--CUE phase shifts. We underline that
the observed COE--CUE phase shift distribution for the quasi--$1d$ conductor
is less trivial than the COE--CUE character of the reflection matrices
for strong quasi--$1d$ localization which only results from the isotropy
assumption. A similar decoupling of the $S$-matrix in two nearly independent
blocks have also been discussed recently by Borgonovi and Guarneri \cite{BG}.

If $Y_{2}(r)$ and $b(k)$ are the two-level cluster function and the two-level
form factor of two independent ensembles, the corresponding functions for the
combined ensemble are given by \cite{Pandey}

\begin{equation}
Y_{2}^{s}(r) = \frac{1}{2} \ {Y}_{2}(r/2) \ ,
\label{eq:Y2s}
\end{equation}

\begin{equation}
b_{s}(k) = b(2k) \ .
\label{eq:bs}
\end{equation}

 This expected crossover situation towards two uncoupled COE--CUE,
for sample lengths of order of the localization length, is indeed
observed in Fig. 3 where we show the Fourier components
$\snM$ with and without magnetic field. For the very long,
weakly disordered samples (R10, diamonds, $AR=30$, $W$=1, $N$=14) the phase
shift distribution (inset) is as homogeneous as for the quasi--$1d$ conductor.
The low harmonics of $Y_{2}$ (small $n$ values of $\snM$)
behave as those of two uncoupled
COE--CUE, reflecting the statistical independence of the short length
trajectories contributing to reflection. Higher order harmonics behave more
like a single COE--CUE, indicating that localization is not sufficient for
decoupling the long trajectories.  For a better understanding of this
effect, one can use the semiclassical picture developed in Appendix
\ref{sec:Semi}, where for relatively small values of $n$, $\snM$ is given
in terms of periodic orbits of the closed sample which hit $n$ times the
vertical limiting hard walls: small values of $n$ are
related with short reflection trajectories which stay close to each of
the samples edges and do not explore the other extreme of the sample
(two decoupled COE--CUE behavior).
On the contrary, for the large values of $n$, the contribution of the
transmission trajectories can not be ignored, and removes the statistical
independence of the two reflection matrices.

Increasing of the system length being time consuming for the numerical
simulations, we can more easily achieve localization
by increasing the strenght of the disorder potential keeping the
geometry fixed. In this case we are able to obtain a more complete decoupling,
extending to higher harmonics the characteristic behavior of two
independent COE--CUE (R4--F4, filled circles, $AR=4$,
$W$=4, $N$=14). Having stronger localization yielded by stronger
disorder, we note an additional effect: the phase shift
density is no longer  uniform. We will discuss this departure from isotropy in
detail in
the next section. For the purpose of the present discussion we only indicate
that we numerically unfold the phase shift
spectra to a rescaled spectra of uniform density. Then, one  can see
in Fig. 3 a rather complete statistical decoupling of the two
sample edges. Let us note also that the small
$n$ behavior of $\snM$ is now a little above what we expect from two uncoupled
COE--CUE, a point which will be considered in Sec. \ref{sec:Unfol}.

In order to study more precisely how the transition from the metallic case
(COE or CUE-like cases) to the localized regime takes place, we calculate the
number $n(r)$ of levels contained in an interval of
length $r$ for the unfolded spectum of the phase shifts and its variance
(number variance)

\begin{equation}
\Sigma^{2}(r) = \langle \left( n(r) - r \right)^2 \rangle \ ,
\label{eq:numvar}
\end{equation}

\nin which can be obtained directly from the numerical data and compared with
universal forms of the standard ensembles. Using Eq. (\ref{allequations}),
one gets for the unitary and orthogonal
cases \cite{Mehta}

\begin{mathletters}
\label{alles}
\begin{equation}
\Sigma^{2}_{UE}(r) = \frac{1}{\pi^2} \left( \ln{(2 \pi r)} + {\cal C} + 1 -
\cos {(2 \pi r)} - {\rm Ci}(2 \pi r) \right) + r \left(1 -
\frac{2}{\pi} {\rm Si}(2 \pi r) \right) \ ,
\label{ea}
\end{equation}
\begin{equation}
\Sigma^{2}_{OE}(r) = 2 \Sigma^{2}_{UE}(r) +
\left(\frac{{\rm Si}(\pi r)}{\pi} \right)^2 - \frac{{\rm Si}(\pi r)}{\pi} \ ,
\label{eb}
\end{equation}
\end{mathletters}

\nin with the large-$r$ behavior

\begin{mathletters}
\label{allels}
\begin{equation}
\Sigma^{2}_{UE}(r) = \frac{1}{\pi^2} \left( \ln{(2 \pi r)} + {\cal C} + 1
\right) + {\cal O}(r^{-1}) \ ,
\label{ela}
\end{equation}
\begin{equation}
\Sigma^{2}_{OE}(r) = \frac{2}{\pi^2} \left( \ln{(2 \pi r)} + {\cal C} + 1 -
\pi^{2}/8 \right) + {\cal O}(r^{-1}) \ .
\label{elb}
\end{equation}
\end{mathletters}

For the superposition of two independent ensembles, one gets\cite{Pandey}

\begin{equation}
\Sigma^{2}_{s}(r) = 2 \ \Sigma^{2}(r/2) \ .
\label{Sigmas}
\end{equation}

Since our original phase shift spectrum is bounded between 0 and $2 \pi$,
$\Sigma^{2}(r)$ folds back to 0 for $r = M$ as the number of phases in
$0 < \theta < 2 \pi$ (or $0 < \Theta < M$) is always $M$. Hence our
comparisons between our numerical data and Eqs. (\ref{alles}) -
(\ref{Sigmas}) are meaningful for $r \ll M$ and we focus our attention
to intervals $0 < r < M/3$.

In Fig. 4 we show how the number variance $\Sigma^{2}(r)$ changes
from the metallic to the localized regime when we increase the disorder or the
sample length. If the disorder is increased, we obtain a continuous transition
from the COE case towards less rigid spectrums and the phase shift density
develops a more and more non uniform structure. The strength of the local
disorder not only decouples
the right and left reflections, but also introduces in the transverse
direction a dimensionality dependent dynamics, which breaks isotropy.
The number variance can then exceed that of two uncoupled COE.

When the disorder is small, and localization is achieved by increasing the
system length, the transverse dynamics remains essentially zero dimensional,
$S$ is isotropic and the number variance indicates a cross-over from
approximately a single COE for the $2N$
phase shifts of $S$ (quasi--$1d$ conductor) to two decoupled COE--like
sets of $N$ phase shifts associated to right and left reflections.

When a magnetic field is applied, we obtain similar results, with a
slightly improved agreement with the CUE--like character: the applied
magnetic field suppressing coherent interferences between time reversed
trajectories shifts the sample towards the metallic regime and doubles
the localization length \cite{Pichard}.

\section {Average Scattering Matrix and Isotropy Hypothesis}
\label{sec:Mean}

 Quasi--$1d$ distributions are partly based on the isotropy hypothesis: i.e.
the unitary matrices $u^{(l)}$ are uncorrelated with the radial part
of $S$ and distributed with the invariant measure on the unitary
group. This implies that the ensemble average of $S$ must be zero.
This property does not hold for high disorder and non quasi--$1d$
geometry. For instance, a non uniformity of the phase shift distribution
was noticed in the previous section for a large disorder and occurs
for weaker disorder in shorter samples (squares and thin slabs). This source
of discrepancy with the circular ensemble is studied in this
section.

In the circular ensembles, the phase shift density is uniform
and equal to $\rho_{0} = M/2\pi$. This means that $\langle Tr S^n\rangle
= 0$ for all $n \neq 0$. We will see that this is not the case for
disordered systems. The first harmonic of the Fourier expansion of
the phase shift density $\rho(\theta)$ is:

\begin{equation}
\rho_{1}(\theta) = \frac{1}{\pi} \ Re[\exp{(- i \theta)}
\langle Tr S \rangle] \ ,
\label{eq:rho1}
\end{equation}

\nin and $\langle Tr S \rangle = 0$ is a necessary condition to obtain
uniform phase shift distribution. Since

\begin{equation}
\langle Tr S \rangle = \sum_{a=1}^{N} ( \langle r_{aa} \rangle
 + \langle r'_{aa} \rangle) \ ,
\label{eq:TrS}
\end{equation}

\nin we can calculate $\rho_{1}(\theta)$ by evaluating the mean values of
the diagonal reflection elements in perturbation theory.

The transmission (reflection) amplitude from a mode $a$ on the left
to a mode $b$ on the right (left) for electrons at the Fermi energy
$E_{f}=\hbar^2 k^2/2m$ is given by \cite{FishLee}

\begin{mathletters}
\label{alltrs}
\begin{equation}
t_{ba}=-i\hbar(v_{a}v_{b})^{1/2}\int dy^{\prime}\int dy \
\phi_{b}^{*}(y^{\prime}) \ \phi_{a}(y) \ G_{k}(L_{x},y^{\prime};0,y)
\label{tra}
\end{equation}
\begin{equation}
r_{ba}=\delta_{a,b}-i\hbar(v_{a}v_{b})^{1/2}\int dy^{\prime}\int dy \
\phi_{b}^{*}(y^{\prime}) \ \phi_{a}(y) \ G_{k}(0,y^{\prime};0,y)
\label{trb}
\end{equation}
\end{mathletters}

\nin where $v_{a}$ ($v_{b}$) and $\phi_{a}$ ($\phi_{b}$) are the
longitudinal velocity and transverse wavefunction for the incoming
(outgoing) mode $a$ ($b$). For hard-wall boundary conditions, the transverse
wave functions have the sinusoidal form presented in Sec. \ref{sec:Scat},
$v_{n}=\hbar k_{n}/m$, $k^2_{n}=k^2-(n \pi/L_{y})^2$, $n=a,b$. We note by $m$
the effective mass of the electrons. For the transmission (reflection)
amplitudes $G_{k}({\bf r'};{\bf r})$ is the retarded Green function evaluated
at the Fermi energy between points ${\bf r} = (x,y)$ on the left lead
and ${\bf r'} = (x',y')$ on the right (left) lead.
Similar expressions hold for the transmission (reflection) amplitudes
for modes comming from the right by using $G_{k}(0,y';L_{x},y)$ in
Eq.~(\ref{tra}) and $G_{k}(L_{x},y';L_{x},y)$ in Eq.~(\ref{trb}) instead of
$G_{k}(L_{x},y';0,y)$ and $G_{k}(0,y';0,y)$ (and placing the $y$ abscissa at
$x=L_{x}$, and $y'$ at $x=0$ ($L_{x}$)).

For a given impurity configuration, the unaveraged retarded Green function
$G_{k}$ for electrons at the Fermi level in the absence of a magnetic field
satisfies

\begin{equation}
\left( \frac{\hbar^2}{2 m} \bigtriangledown^{2}_{\bf r} +
\frac{\hbar^2 k^2}{2 m} + V({\bf r}) + i \gamma \right) G_{k}
({\bf r'};{\bf r}) = \delta ({\bf r'} - {\bf r}) \ ,
\label{eq:GR}
\end{equation}

\nin with $\gamma \rightarrow 0^{+}$. We will assume that the impurity
potential $V({\bf r})$ is given by $N_{i}$
uncorrelated $\delta$-function scatterers (of strength $u$) randomly
distributed in the disordered strip.

\begin{equation}
V({\bf r}) = \sum_{\alpha=1}^{N_{i}} u \ \delta ({\bf r} - {\bf R}_{\alpha})
\hspace{1cm} \left\{ \begin{array}{l}
0 < R_{\alpha,x} < L_{x} \\
0 < R_{\alpha,y} < L_{y}
\end{array} \right. \hspace{1cm} \alpha=1,\ldots,N_{i} \ .
\label{eq:pot}
\end{equation}

The standard techinque used in disordered systems is to solve Eq.
(\ref{eq:GR}) in perturbation theory and take the ensemble average at
each order of the perturbation expansion \cite{Feng,Akk}.
As indicated diagramatically in Fig. 5,  we merely expand to second order
in the perturbation expansion.

The unperturbed Green function $G_{k}^{(0)}({\bf r'};{\bf r})$ for an infinite
strip can be expanded in the base of the transverse wavefunctions
$\phi_{n}$, and the $n^{\rm th}$ coefficient is a one-dimensional Green
function with an effective wave vector
$k_{n} = \sqrt{k^2 - (n \pi/L_{y})^2}$ ($k_{n}$ is real when $n$ corresponds
to a propagating mode and pure imaginary otherwise),

\begin{equation}
G_{k}^{(0)}(x',y';x,y) = \frac{m}{i \hbar^2} \sum_{n=1}^{\infty}
\frac{1}{k_{n}} \exp{(i k_{n} |x'-x|)} \phi_{n}(y') \phi_{n}(y) \ .
\label{eq:G0}
\end{equation}

The inclusion of $G_{k}^{(0)}$ in (\ref{trb}) just cancels the $\delta_{a,b}$
factor since in the absence of disorder the modes propagate without any
reflection. The first and second order corrections are respectively given by

\begin{mathletters}
\label{allr12s}
\begin{equation}
\langle r_{ba}^{(1)} \rangle = \langle r_{ba}^{\prime (1)} \rangle =
- \delta_{a,b} \ \left(\frac{m u}{2 \hbar^2}\right)
\frac{n_{i}}{k^2_{a}} \ (1 - e^{2 i k_{a} L_{x}}) \ ,
\label{r12a}
\end{equation}
\begin{equation}
\langle r_{ba}^{(2)} \rangle = \langle r_{ba}^{\prime (2)} \rangle =
i \ \delta_{a,b} \ \left(\frac{m u}{2 \hbar^2}\right)^2
\frac{n_{i}}{k^2_{a}} \ (1 - e^{2 i k_{a} L_{x}}) \ ,
\label{r12b}
\end{equation}
\end{mathletters}

\nin where $n_{i}=N_{i}/(L_{x}L_{y})$ is the impurity density. For the
second order correction we are only giving the leading term (in the
impurity parameter $u n{_i}$, and in the inverse mode number $1/N$),
and we have cut the sum over
the internal momentum (as usually done for $\delta$-function potentials).
The highly oscillating phases $e^{2 i k_{a} L_{x}}$ would be suppressed
by inelastic scattering.
For translationally invariant (after averaging) systems the perturbation
theory is usually done in momentum representation; the
first order term gives rise to a real self-energy that merely renormalizes
the Fermi energy, while the second order term gives rise to an imaginary
self energy which is responsible for an exponential damping of the average real
space Green function. For the reflection amplitudes from finite disordered
regions, we take into account perturbation up to second order term only.
This calculation neglects the multiple scattering processes characteristic
of the diffusive regime. However, we are only interested in the average
value of the reflection amplitude, where single scattering dominates multiple
scattering, as we will see in our semiclassical approach, and
Eqs.~(\ref{allr12s}) give the main features seen in the numerical simulations.
They are proportional to the strength of the disorder and show a non-zero
average
only for the diagonal elements of the $S$-matrix. The denominator
$k_{a}$ indicates that the absence of self-averaging is more pronounced
for the higher modes.  The breakdown of perturbation
theory for small $k_{a}$ is understandable since it is close to
a threshold of complete reflection which cannot be obtained perturbatively.
Clearly, those threshold effects, when new conduction channels appear
($N \rightarrow N+1)$), drastically limit the validity of the
isotropy assumption.

Up to second order perturbation the average diagonal reflection amplitudes
are

\begin{equation}
|\langle r_{aa}^{(1,2)} \rangle | = |\alpha| \ \frac{n_{i}}{k^2_{a}}
\ \sqrt{2(1 - \cos{(2 k_{a} L_{x})})} \hspace{0.8cm} ,
\label{eq:raa12}
\end{equation}

\nin where $\alpha = \left(\frac{m u}{2 \hbar^2}\right) \left(-1 + i
\left(\frac{m u}{2 \hbar^2}\right)\right)$. For the lowest modes Eq.
(\ref{eq:raa12}) gives a correction vanishing as $1/k^2$ (or $(L_{y}/N)^2$),
but the correction remains important for the highest modes. On the other
hand, in our numerical simulations in a lattice model, $N$ will always be
finite and not very large. In Fig. 6.a we show the values of
$| \langle r_{aa} \rangle |$ obtained from our numerical simulations, as a
function of the mode number $a$. The average reflection amplitudes of the
sample described in the Sec. \ref{sec:metal1}
(R1, squares, $AR=4$, $W$=1, $N$=14) are close to the functional form
$1/k_{a}$ (solid thick line) and to those of a longer sample (R8, circles,
$AR=10$, $W$=1, $N$=14). The average value of the nondiagonal
reflection amplitudes (not shown) are zero within the statistical error.
Since this perturbation calculation is performed in the continuum and yields a
rapidly oscillating phase we do not expect to get full agreement with the
numerical simulations on a lattice. We are mainly interested in the mean
behavior of the diagonal reflection elements as a function of the channel
number $a$, the number of modes $N$ and the strenght of the disorder $W$.
Increasing the disorder (R2, filled diamonds, $AR=4$, $W=1.5$, $N=14$)
enhances the correction. Keeping the number of modes $N=14$ (Fig. 6.b) while
going to a square geometry (R12, squares, $AR=1$, $W=1$, $N=14$) decreases the
Fermi momentum and gives a larger correction for the average reflection
amplitude than for the sample of Sec. \ref{sec:metal1}. The correction for a
square
geometry is of the same order of that of a long sample (R11, circles, $AR=4$,
$W=1$, $N=14$) with the same number of modes and cross section (same Fermi
momentum).
Increasing the disorder (R13, diamonds, $AR=1$, $W=1.5$, $N=14$) enhances the
correction, while increasing the number of modes (R18, full triangles,
$AR=1$, $W=1.5$, $N=28$) decreases the correction.

The mean values of the scattering matrix elements have been considered by
Iida, Weidenm\"uller and Zuk \cite{IWZ} who studied the interplay between
universal conductance fluctuations and the statistical properties of the
hamiltonian spectrum of a disordered strip. They define the sticking
probability $R_{a} = 1 - | \langle r_{aa} \rangle |^2$, measuring the
weight of the fast processes (where the particle is reflected after a few
scattering events in the disordered region)
versus the long trajectories. Sticking
probabilities smaller than 1 give corrections to the universal quasi-1D
value of the variance of the conductance fluctuations, which can be calculated
with the aid of Eq. (\ref{eq:raa12}).

This non-zero average of the diagonal refection elements is in disagreement
with the standard maximum entropy approaches where one assumes that the
matrices $u^{(l)}$  are uniformely
distributed in the unitary group, giving zero average
values of the $S$-matrix. Recent work by Mello and Tomsovic \cite{MT} has
relaxed the isotropy assumption, making possible a non-zero mean for
the transmission amplitude (diagonal in mode number and exponential in
$l/L_{x}$). Our results show that a non-zero average for the diagonal
reflection amplitudes is needed in order to describe systems outside the weak
disorder and quasi-one-dimension.

In the large $N$-limit we approximate the second order perturbation calculation
of $\langle Tr S \rangle$ by converting
the sum over modes into an integral, the phases $\exp{(2 i k_{a} L_{x})}$
give rise to higher order terms in $1/N$, and from Eqs. (\ref{allr12s})
we obtain (away from the thresholds)

\begin{equation}
\langle Tr S^{(1,2)} \rangle \approx \alpha \ \frac{n_{i} L_{y}}{k \pi}
\ \ln{\left( \frac{k+N\pi/L_{y}}{k-N\pi/L_{y}} \right)}
\approx \alpha \ \frac{n_{i} L_{y}^2}{\pi^2}
\ \frac{1}{N} \ \ln{\left( 2 N\right)} \ .
\label{eq:Trpert}
\end{equation}

It is important to notice that Eq. (\ref{eq:Trpert}) does not depend on the
length $L_{x}$ of the sample (or the aspect ratio) but only depends on the
properties at the entrance of the sample (transverse cross-section $L_{y}$
and Fermi momentum $k$). Its logarithmic dependence on $N$ indicates that
there is homogeneity in the phase distribution only for very large $N$ and
very small transverse cross section.

 From Eqs. (\ref{eq:rho1}) and (\ref{eq:Trpert}) we can see that the phase
shift distribution will not be uniform, unless we have a narrow sample
with weak disorder and a large number of propagating modes. A non uniform
density is often the case for hamiltonian spectra (or for
the radial parameters of $M$), and one studies the correlations of the
unfolded spectra\cite{RevBoh,SPM}. Instead of studying the bare phase
shifts $\theta_{m}$, we have to consider the rescaled variable
$\Theta_{m} = {\cal N}(\theta_{m})$,
where ${\cal N}(\theta)$ is the number of levels below $\theta$, i.e.
${\cal N}(\theta) = \int_{0}^{\theta} \rho(\theta^{\prime}) d \theta^{\prime}$.
Since our interval of phase shifts is bounded between 0 and $2 \pi$, for
the numerical study, we repeat twice each phase shift sequence and
calculate the two-point correlation function in the
second interval ($M$ to $2 M$ for the unfolded variables $\Theta_{m}$).
The Fourier components $\snM$ can then be compared with the universal
two-level form factors of Eq. (\ref{allbes}), as we do in the following
sections.

\section {Unfolded Phase Shift Spectra Outside Quasi-One-Dimension}
\label{sec:Unfol}

We now turn our attention to samples far from a quasi-$1d$ shape.
For disordered squares (Fig.~7.a) the phase
shift densities are non uniform in the absence of
magnetic field (inset). The degree of nonuniformity is consistent with
Eqs. (\ref{eq:rho1}) and (\ref{eq:Trpert}), i.e. it becomes more important
for higher disorder and smaller number of propagating channels. The Fourier
components $\snM$ (after unfolding) are in relatively good agreement with
the COE--CUE two-level form factors, but the correspondence becomes poorer
when we increase the disorder. For the large-$n$ Fourier
components, we are probing the correlations for small separations,
where we can take the phase shift density as constant. This approximate
translational invariance makes that only the term with $n = - n'$ survives
in the expansion of the delta functions of Eq. (\ref{eq:R2def}).
The Fourier
components of the two-point correlation function are still given by
$\langle |TrS^n|^2 \rangle$, through Eq. (\ref{sn}).
For squares with magnetic field and low disorder the phase shift density is
relatively uniform (lower histogram of the inset) and we use Eq. (\ref{sn})
instead of unfolding the spectrum.

In Fig.~7.b, a disordered slab is considered
(upper inset, the incoming electron flux is along the direction of the
shortest dimension). The phase shift distributions (inset) are strongly non
uniform ($L_{y}$ in Eq. (\ref{eq:Trpert}) is very large), but the unfolded
spectra, with (triangles) and without (filled squares) magnetic field, are
relatively well described by Eq. (\ref{allbes}), except
for small $n$ values where a careful look indicates values above the COE--CUE
behavior. This is another important source of departure from the universal
behavior related to transverse diffusion, in
addition to the cross--over mentioned in section \ref{sec:isol1} coming
from longitudinal localization.

The approximate agreement of the numerical
data with the two-point form factors of the circular ensembles is
somewhat surprising in these geometries.
Given the approximate agreement of the unfolded spectra with the standard
ensembles at the level of the two-point correlation function, we might ask
at this stage whether the unfolded phase shift
spectrum of metallic conductors is well described by the circular
ensembles, independently of the shape and strenght of the disorder.
However, a check at the level of the two-point correlation function
is not very accurate, as we have learned from statistical studies
of chaotic hamiltonians \cite{RevBoh} and transmission
matrices \cite{Sle}. A better test is provided by integrals involving
the two-point correlation function. As in Sec. \ref{sec:isol1},
we now consider the number statistics $n(r)$ and the number
variance (Eq. (\ref{eq:numvar}))

In order to analyze systematic departures from the random matrix
correlations
we plot in Fig. 8, for various geometries, degrees of disorder and
number of propagating modes, the difference of the number variances between
the numerical data and the COE--CUE values (\ref{alles}). We consider
$\sigma^{2}(r) =
\Sigma^{2}(r) - \Sigma^{2}_{OE}(r)$ for samples without magnetic field and
$\sigma^{2}(r) = \Sigma^{2}(r) - \Sigma^{2}_{UE}(r)$ for samples with
nonzero magnetic field. Like in the previous section, we only show the interval
$0 < r < M/3$ where the comparission is meaningful.
One can see very clearly now that, even after
unfolding, the accuracy of the random matrix description
strongly depends on the shape and degree of disorder of the
samples. The difference $\sigma^{2}(r)$ in number variances grows
approximate linearly with $r$. The magnitude of the slope of
$\sigma^{2}(r)$ measures the validity of the random matrix
description. Squares with low disorder (after unfolding) are well
represented by the standard ensembles, the agreement becomes
poorer when increasing the disorder, and improves when augmenting
the number of modes $N$. However, slab-shaped samples show also
large deviations respect to the COE--CUE values. The conditions
for having a good COE--CUE distribution coincide with the conditions
for diminishing
$|\langle Tr S \rangle|$ and obtaining a more uniform distribution
(Eqs. (\ref{eq:rho1}) and (\ref{eq:Trpert})). Notice however that Fig. 8 is
done over unfolded
spectra, where the nonhomogeneous phase shift distribution is in principle
already accounted for. Our task in the remaining section is to
to quantitatively study the deviations to the random matrix behaviors in
disordered conductors, giving the slope of the curves $\sigma^{2}(r)$
as a function of the aspect ratio, degree of disorder and number of
propagating modes of the disordered sample.

\section {Deviations From The Universal RMT Behavior}
\label{sec:Devi}

The statistical properties of the spectra of small disordered systems have been
considerably more studied than those of the scattering phase shifts.
It is then appropriate to establish a connection between the
two in order to understand our numerical results. The relationship between
phase shifts and energy levels has been investigated in the
semiclassical limit by Bogomolny\cite{Bog} and by Doron and
Smilansky\cite{DoSm}. Their approach is not completely applicable to our case
but provides some guiding concepts. We reproduce in Appendix
\ref{sec:Rela} their main results and
discuss the points where the correspondence does not hold.

Under the assumptions discussed in Appendix \ref{sec:Rela} of neglecting the
evanescent modes and the nonuniformity of the phase shift distribution,
the energy level density is given by

\begin{equation}
{\rm d}(E) =
\frac{M}{2\pi\hbar} \langle \tw(E)\rangle \ ,
\label{denseigen2}
\end{equation}

\nin where $\tw(E)$ is the Wigner time

\begin{equation}
\tw(E) =  \frac{\hbar}{iM} Tr\left(S^{\dagger}(E) \frac{dS(E)}{dE}
\right) \ ,
\label{tauw}
\end{equation}

\nin whose physical interpretation is discussed in Appendices \ref{sec:Semi}
and
\ref{sec:Rela}. Assuming
that the density of states of our disordered rectangular samples is the same as
without disordered ${\rm d}(E) = 1/\Delta = mL_{x}L_{y}/2\pi\hbar^{2}$, we can
check the validity of Eq. (\ref{denseigen2})) against our numerical
simulations. As indicated in Table I of Appendix \ref{sec:Summa} there is
agreement within 5 to 25 \% in most of the cases, which is reasonable
given the various approximations involved.

The quantity most often calculated in metallic spectra is
the density-density correlation function

\begin{eqnarray}
K_{2}(E,E+\varepsilon) = \sum_{n,n'} \langle \delta (E -E_{n})
\delta (E+\varepsilon -E_{n'}) \rangle \ - \ {\rm d}(E) \ {\rm
d}(E+\varepsilon)
\label{K2}
\end{eqnarray}

Scaling the energy separation with the mean level spacing ($e=
\varepsilon {\rm d}(E)$) and the phase shift separation with the mean
phase shift distance ($r = \eta M/2\pi$) we have, from Eq.~(\ref{K2b}),
in the large $M$ limit that

\begin{equation}
K_{2}(e/{\rm d}(E)){\rm d}^{-2}(E) = \delta(r)-Y_{2}(r) \ ,
\label{K2c}
\end{equation}

As stated before, we will
put aside the fact that the two above mentioned
assumptions are not quite true in our
systems and we will pursue the consequences  of Eq. (\ref{K2c}).

The density-density correlation function for disordered systems
has been obtained in perturbation theory by
Altshuler and Shklovski\u{\i}
\cite{AS}

\begin{equation}
K_{2}(\varepsilon) = - \frac{s^2}{\pi^2}
Re \sum_{\{n_{\mu}\}} (\varepsilon + i\hbar D q^2 + i\gamma)^{-2} \ ,
\label{eq:AS}
\end{equation}

\nin for energies $\varepsilon$ large compared to the level spacing $\Delta$,
and small compared with the energy scale $\hbar / \tau_e$, associated with the
elastic scattering time $\tau_e$. The factor $s$ accounts for the spin
degeneracy of each level ($s=1$ since we work with spinless electrons), and
$\gamma$ is a small energy cutoff (to account for
inelastic scattering). For simplicity we will be work the case of zero
magnetic field. The sum is over the diffusion modes in the sample,
assumed to be a $d$-dimensional parallelepiped with sides
$L_{\mu}$, that is, $q^2=\pi^2 \sum_{\mu=1}^{d}(n_{\mu}/L_{\mu})^2$.
The diffusion coefficient is $D=v_{f} l/d$. The Thouless energy
$E_{c,\mu}=\hbar D/L_{\mu}^2$
is inversely proportional to the time that takes an electron to diffuse across
the sample in the $\mu$-direction. For samples with all $L_{\mu}$ equal
(hypercubes) we just have one Thouless energy. That will be the case of our
square samples, while for quasi-one-dimensional samples we have in principle
two Thouless energies, but we will reserve this name for the smaller one, that
is, the one associated with the length $L_{x}$.

The mean square fluctuation in the number of levels (number variance) is given
in terms of the density-density correlation function,

\begin{equation}
\langle[\delta N(\varepsilon)]^{2} \rangle =
\int_{E-\varepsilon/2}^{E+\varepsilon/2} \ d E_{1}
\int_{E-\varepsilon/2}^{E+\epsilon/2} \ d E_{2} \ K_{2}(E_{1},E_{2}) \ ,
\label{eq:msfnl}
\end{equation}

\nin which from (\ref{eq:AS}) can be written as

\begin{equation}
\langle[\delta N(\varepsilon)]^2\rangle = \frac{1}{\pi^2} \sum_{\{n_{\mu}\}}
\ln{\left(\frac{\varepsilon^{2}}{(\gamma + \hbar D q^2)^{2}} + 1 \right) } \ .
\label{eq:msfnseq}
\end{equation}

For energies $\varepsilon \ll E_{c,\mu}$ the sum (\ref{eq:msfnseq}) is
dominated
by the term with all $n_{\mu}$ null and \cite{AS}

\begin{equation}
\langle[\delta N(\varepsilon)]^2\rangle = \frac{1}{\pi^2}
\ln{\left(\frac{\varepsilon^{2}}{\gamma^{2}} + 1 \right)} \ .
\label{eq:msfnsel}
\end{equation}

The perturbation theory of Ref. \cite{AS} is valid for energy separations
$\varepsilon$ larger than the inelastic scattering $\gamma$ or the level
spacing $\Delta$. In our case there is no inelastic scattering and we
substitute $\gamma$ by $\Delta$ obtaining the standard random matrix theory
result $\langle[\delta N(\varepsilon)]^2\rangle = 2/\pi^2
\ln{(\varepsilon/\Delta)}$ for $\Delta \ll \varepsilon \ll E_{c}$.
For energies $\varepsilon \gg E_{c}$ the summation over $\{n_{\mu}\}$ can be
replaced by an integral over $dq_{\mu}$ and

\begin{equation}
\langle[\delta N(\varepsilon)]^2\rangle = c_{d} \left(\frac{\varepsilon}
{E_{c}} \right)^{d/2} \ ,
\label{eq:msfnseg}
\end{equation}

\nin where $c_{d}$ is a dimensionality-dependent numerical coefficient
\cite{AS}. The asymptotic results (\ref{eq:msfnsel}) and (\ref{eq:msfnseg})
have recently been rederived by Argaman, Imry and Smilansky \cite{AIS} using
a more intuitive semiclassical method for electrons in the diffusive regime.
The small-energy universal regime is obtained for times long enough to allow a
diffusing electron to ergodically explore all the sample, while for energy
intervals larger than the Thouless energy
(short times) the correlations depend on the diffusing,
unbounded dynamics of the electron, which is dimensionality and disorder
dependent. The universal character of the short range
eigenvalue correlation and the long range non-universal (dimensionality
and disorder dependent) part have been also obtained in numerical simulations
of
by Dupuis and Montambaux \cite{Monta} who studied the crossover between the
two regimes in an Anderson model.

Given Eq. (\ref{K2c}), and the above results for the number variance
predicted by perturbation theory, we expect the correlation functions
of the
phase shifts to have a COE--CUE behavior for phase shift separations
smaller than

\begin{equation}
\eta_{c} = \frac{\tw E_{c}}{\hbar} \ .
\label{etac}
\end{equation}

Our numerical data summarized in Table I of Appendix A indicates
that for the studied samples the critical angle
$\eta_{c}$ is smaller than the mean phase shift spacing ($r_{c} =
\eta_{c}/(2\pi/M) = \langle g \rangle /2\pi < 1$).
Therefore they are in the transition regime between the two
asymptotic limits (\ref{eq:msfnsel}) and (\ref{eq:msfnseg}).

In Secs. \ref{sec:isol1} and \ref{sec:Unfol} we have studied the number
variance of the phase shifts
$\sigma^{2}(r) = \Sigma^{2}(r) - \Sigma^{2}_{OE}(r)$. In particular,
from Fig. 8 we can see that $\sigma^{2}(r)$ grows almost linearly from
$r \approx r_{c}$. To check our prediction, it is interesting to
look at the value of the slope.
Whithin the perturbation theory of Ref. \cite{AS} the difference between the
number variance for the real energy levels and the R.M.T variance
is obtained by excluding the term $n_{x}=n_{y}=0$ from the sum
(\ref{eq:msfnseq}). For a two-dimensional sample the slope of the
difference evaluated at $\nu_{c}=E_{c}/\Delta$ is given by

\begin{equation}
\left( \frac{d}{d e} \langle[\delta N(e)]^2\rangle
- \Sigma^{2}_{OE}(e) \right) \
= \frac{2}{\pi^2} \frac{\Delta}{E_{c}}
\sum_{\{n_{x},n_{y}\}\neq \{0,0\}} \frac{1}{1+\pi^4(n_{x}^2+n_{y}^2)^2}
\approx 0.0214 \frac{1}{e_{c}} \ .
\label{eq:slope}
\end{equation}

\nin For a quasi-one-dimensional sample the sum is over $n_{x} \neq 0$ and the
slope is approximately $0.0045/e_{c}$. Given the relationship
between the statistics of phase shifts and energy levels, we expect
that the
slope $\phi_{c} = \frac{d}{dr}\sigma^{2}(r)|_{r=r_{c}}$ also scales linearly
with $1/r_{c}$. In Fig. 9 we show the value
$\phi_{c}$ versus the inverse Thouless energy
$1/r_{c}$.  The linear relationship predicted by Eq.
(\ref{eq:slope}) turns out to be approximately valid. The slope obtained for
the two-dimensional case (squares geometries in Fig. 9) agrees
within $50 \%$
with the coefficient of (\ref{eq:slope}) and is a factor of 4 larger than the
slope obtained for quasi-one dimensional geometries (roughly the same ratio
than for the eigenenergies). Obviously, in those samples,
the quasi-one-dimensional limit is not achieved and the approach to this
limit depends on the aspect ratio of the sample.

\section {Conclusions}
\label{sec:Concl}

In this work we have considered scattering in disordered systems.
We calculated the
invariant measure of $S$ in the polar decomposition and compared the
COE--CUE distributions with those of the global and
local approaches to the transfer matrix of disordered systems. We then
turned to the study of the scattering phase shifts for different
geometries
and degree of disorder. For quasi-one dimensional samples in the metallic
regime the density and the two-point correlation functions are
close to those of the circular ensembles. Increasing the lenght with
fixed disorder or
the disorder with fixed geometry, we break $S$
into two uncorrelated reflection submatrices. Deep in the quasi--$1d$
localized regime the decoupling
is almost complete and the two-point correlation function of the phase
shifts is that of a superposition of two independent circular ensembles
(COE or CUE).

Outside quasi-one dimension, the phase shift
density differs from the uniform distribution
characteristic of Dyson ensembles.
This anisotropy comes from short-time processes which yield a non vanishing
ensemble average of $S$, and is strongly enhanced in the vicinity of a
energy threshold where a new conduction channel appears.
Using diagrammatic perturbation theory, we
calculated the average value of the reflection amplitudes, finding good
agreement with the numerical simulations and establishing the conditions
for having the uniform phase shift distribution of the circular ensembles.
One needs both energy far from  a threshold, weak-disorder,
small transverse length and large number of
propagating modes. When this is not the case, anisotropy is important and
we have to unfold the spectrum in order to study its correlations.

For large transverse lenghts the electron diffusion in the perpendicular
direction gives rise to small-$k$ corrections of the
two-level from factor $b(k)$ of the [unfolded] phase shift spectrum. This
non universal behavior is related to the large-energy (non-ergodic)
behavior found by Altshuler and Shklovski\u{\i} for the energy-level
statistics of small metallic particles. This lead us to relate under
certain assumptions the scattering phase shift and the energy-level
form factors.
We numerically checked this relationship for the number variance (fluctuation
of the number of states within a given interval). We verified that the
non universal discrepancies become increasingly important as the Thouless
energy decreases, in a geometry dependent fashion.

One of the most important problems remaining in the random matrix theory
of electron transport in disorder systems is the extension of the standard
approach outside the quasi-one-dimensional case. We have shown in this work
that anisotropy is an essential ingredient in higher dimensions.
The precise form of the decoupling
of the $S$ matrix into independent reflections blocks in the one-dimensional
localized regime is another interesting problem left for futures studies.
Given the recent progress in the theory of parametric correlations
\cite{Aaron,SiAl,Been3},
the energy dependence of the scattering phase shifts desserves
further studies (e. g. the correlations of the
characteristic times $\tau_{m}(E)$ (Eq.~(\ref{taum}), slopes in Fig. 10)
which we are presently developing.

\section {Acknowledgements}
\label{sec:Ackno}

We acknowledge helpful discussions with B.~Altshuler, H.~Baranger,
C.~Beenakker,
E.~Bogomolny, S.~Feng, P.~Leboeuf, P.~Mello, G.~Montambaux,
K.~Slevin, U.~Smilansky and H. Weidenmuller. Both authors thank
the Wissenschaftskolleg zu
Berlin for hospitality and fruitful discussions with the participants of the
program on Dynamical Systems. This work was partially supported by EEC,
Contract No SCC-CT 90-0020.

\appendix
\section {Summary of the numerical simulations}
\label{sec:Summa}

In this appendix we summarize the results of the numerical simulations and we
briefly indicate the way they were obtained. The simulations were performed in
a tight-binding model with a number of sites $L_{x} \times L_{y}$ and a random
on-site disorder of amplitude $W$ by using a recursive Geen function method
\cite{LF,BDJS}. In a magnetic field we use the Peierls substitution to relate
the hopping matrix element to the vector potential. The magnetic field is
taken to be linearly increasing in the leads from zero to its full value in
the disordered region over a distance of the order of the transverse dimension
$L_{y}$. For the small fields that we work with the statistical results are
independent on the way the field is introduced.

Runs 1-11 in Table I are in the quasi-one-dimensional limit, while runs 12-18
are for square samples and runs 19 for thin slabs. The time limitation for
the numerical simulations is the transverse size and that limits the number
of samples ($N\!S$) considered for large $L_{y}$. The number of modes $N$ was
kept low (14-20 in most of the cases) in order to have small Fermi energy
($E_{f}$) and wavevector ($k$) avoiding lattice
effects. The Wigner time $\tw$ is calculated from Eq.~(\protect\ref{tauw}) and
provides, through (\protect\ref{denseigen2}), the value of the mean level
spacing. Using the free-space two-dimensional density of states to give the
Weyl term of our disorder region, the level spacing is
$\Delta=mL_{x}L_{y}/2\pi\hbar^{2}$, and the column showing
$(\tw\Delta/\hbar) \ (N/\pi)$ checks the approximate validity of
Eq. (\protect\ref{denseigen2}) under the hypothesis that the level spacing does
not change by the effect of confinement or the disorder. We obtain in most of
the cases an agreement within 5-25 \% is quite
reasonable given the various approximations involved.

The elastic-mean-free path $l$ calculated in Born approximation \cite{Har}
using samples of transverse dimension $L_{y}$ and a few lattice sites on the
longitudinal dimension. In the metallic regime it agrees very well with the
mean free path obtained from the Drude formula for the dimensionless
conductance $\langle g \rangle = Nl\pi/2L_{x}$. The Thouless energy in
Anderson units is given by $E_{c}=l/L_{x}^{2} \ \sin{k}$. The reduced
critical angle $r_{c}$ is obtained by scaling the critical angle
$\eta_{c} = \tw E_{c}/\hbar$ (Eq.~(\ref{etac})) with the mean level spacing
$2\pi/M$. The factor $\phi_{c}$ measuring the discrepancy with the standard
ensembles is obtained from Fig. 8 by taking the slope of $\sigma^{2}(r)$
between $r=0.7$ and $r=7$. This is not completely equivalent to the slope at
$r_{c}$ since $r_{c}$ and the range of linear behavior of $\sigma^{2}$
varies from sample to sample, but has the advantage of providing a consistent
definition for all the cases.

Conductances for individual samples are obtained from Landauer's formula,
Eq.~(\ref{eq:Land}), and then averaged over impurity realizations yielding
the average $\langle g \rangle$ and the variance $\langle \delta g^2 \rangle$.
According to the Thouless formula $\langle g \rangle = 2\pi E_{c}/\Delta$ in
the metallic regime, and by using the approximate $\Delta$ of the perfect
square we obtain an agreement of 25\% with Landauer's formula. It is important
to notice that $r_{c}$ is obtained from $l$ and $\tw$, directly extracted
from the simulations (without any assumption over the form of the conductance).
In the localized regime (runs 4,5 and 10) the average conductance
and its variance are dominated by special configurations with anomalously
large values. Therefore averaging over $\log$ is the meaningfull
\cite{Pichard} thing to do. In Table I we give as $\langle g \rangle$ for
these runs the typical conductance $\exp{\langle \log {g} \rangle}$ and
$\langle \delta (\log {g})^{2} \rangle$ in the column of the variances.

In the metallic regime the difference between the average conductance without
and with magnetic field is the weak locatization correction. Mello and
Stone\cite{MS} have calculated
this correction in the quasi-1d limit by diagrammatic and random matrix
approaches and found a value of $1/3$, which is somehow larger than the one
obtained from Table I. It is rather difficult to reproduce the theoretical
value in numerical simulations on disordered strips since getting to the regime
of validity of the theory (diffusive, quasi-1D, metallic and magnetic field
large enough to kill the time reversal symmetry but small enough for not to
alter the classical electron paths) requires extremely large sample sizes
\cite{HarPriv}. The values of the conductance fluctuations
$\langle \delta g^2 \rangle$ for zero magnetic field in the metallic regime
are in good agreement with the diagrammatic values \cite{AUCF,lee1} and
previous numerical simulations \cite{RevWebb,Yeya}. The halfing of the
variance by the effect of the magnetic field is not quite obtained for
$\phi/\phi_{0}=2$, but for ratios of the order of 5-10 the factor of 2 it is
always found. This is somehow surprising since for $\phi/\phi_{0}=2$ we have
already a complete transition towards the unitary ensemble (see Fig. 2).

The mean values of $|\langle Tr S \rangle|$ are in qualitative agreement with
the perturbation result of Eq. (\ref{eq:Trpert}), increasing with disorder
and decreasing with the number of modes $N$.

\section {Relation between the scattering phase shift
 and energy level statistics}
\label{sec:Rela}

 In this appendix we present,
following
Bogomolny\cite{Bog} and Doron and Smilansky\cite{DoSm},
a relation between the statistical
correlations of the phase shifts and the energy levels
in a form
appropriate to the scattering which we consider.

Close to the entrance cross sections
($x \leq 0$ and $x \geq L_{x}$) the
scattering wave functions are given by the asymptotic expressions
(\ref{allwfs}) plus the contribution of the evanescent modes:

\begin{mathletters}
\label{allwfwes}
\begin{equation}
\Psi_{I}(x,y) = \sum_{n=1}^{N} \frac{1}{k_{n}^{1/2}} \left(A_{n} e^{i k_{n} x}
+ B_{n} e^{-i k_{n} x}\right)\phi_{n}(y) + \sum_{n=N+1}^{\infty}
\frac{1}{|k_{n}|^{1/2}} F_{n} e^{|k_{n}| x}\phi_{n}(y) \ ,
\label{wfwea}
\end{equation}
\begin{equation}
\Psi_{II}(x,y) = \sum_{n=1}^{N} \frac{1}{k_{n}^{1/2}} \left(C_{n}
e^{i k_{n}(x-L_{x})}
+ D_{n} e^{-i k_{n}(x-L_{x})}\right)\phi_{n}(y) + \sum_{n=N+1}^{\infty}
\frac{1}{|k_{n}|^{1/2}} G_{n} e^{-|k_{n}|(x-L_{x})}\phi_{n}(y)\ .
\label{wfweb}
\end{equation}
\end{mathletters}

 In Ref. \cite{DoSm}, the contribution of the
evanescent modes for the evaluation of $\Psi_{I}(x=0,y)$ and
$\Psi_{II}(x=L_{x},y)$ is neglected in the semiclassical limit.
This sharp cut off at
$N+1$ in the sum over $n$ is clearly not exact in the case of our
numerical
simulations, where $N$ is typically 14 and higher transverse harmonics are
needed in order to represent the scattering wavefunction at the entrance.
Neglecting the evanescent channels, the scattering wave function
vanishes at the
boundaries ($x=0$ and $x=L_{x}$) if $B = - A$ and $C = -D$, implying that
the
scattering wave function coincides inside the sample  with an
eigenfunction of the hamiltonian (assuming hard-wall
boundary conditions). Therefore, an energy $E$ is
an eigenvalue of the closed system whenever $S(E)$
has an eigenvalue $-1$ \cite{DoSm}

\begin{equation}
{\rm det} (I+S(E)) = 0 \ .
\label{quantcond}
\end{equation}

\nin
As illustrated in Fig. 10, the energy levels can
be obtained by the intersection of the curves $\theta_{m}(E)$ with horizontal
lines at $\theta=(2n+1)\pi$.
The accuracy of this quantization condition
has been recently tested numerically
\cite{SchSm} in the Sinai billiard. Away from the mode thresholds, the errors
in
the
determination of the energy levels are of the order of only a few percent, and
they are
greatly reduced when a few closed channels are kept. On the other hand,
we are not interested in the precise energy levels but in the
relationships
between their statistics and that of the phase shifts. Towards this
end,
we have calculated the nearest neighbour distribution of the energy levels
determined from
(\protect\ref{quantcond})
and found, in the metallic regime,
good agreement with the corresponding Wigner
distributions.
 From the above quantization condition
(\protect\ref{quantcond})
\footnote{If we use the convention
${\hat S} = \left( \begin{array}{lr}
t'	& \hspace{0.5cm} r	\\
r	& \hspace{0.5cm} t
\end{array} \right)$
and we ignore the evanescent modes, the condition to have an
eigenenergy of the closed system with periodic boundary
conditions is that ${\hat S}$ has the eigenvalue 1, that is,
${\rm det}(I-{\hat S}(E))=0$. It is easy to see that ${\hat S}=SU$, with
$U = \left( \begin{array}{lr}
0	& \hspace{0.5cm} I	\\
I	& \hspace{0.5cm} 0
\end{array} \right)$. Therefore, for Dyson Circular Ensembles where the
probability distribution is given by the invariant measure in the
appropriate space,
we do not expect a difference between the statistical properties
of $S$ and ${\hat S}$.}
the level density can be expressed as

\begin{equation}
{\rm d}(E) = \sum_{n} \langle \delta (E - E_{n}) \rangle =
\frac{1}{2\pi\hbar} \sum_{m=1}^{M} \sum_{n=-\infty}^{\infty}
(-1)^{n} \langle \tau_{m}(E) \exp{(i n \theta_{m}(E))} \rangle \ ,
\label{denseigen}
\end{equation}

\nin where the time associated with the $m^{\rm th}$ phase shift is

\begin{equation}
\tau_{m}(E) = \hbar \ \frac{d\theta_{m}(E)}{dE} \ .
\label{taum}
\end{equation}

Ignoring correlations between $\theta_{m}$ and $\tau_{m}$ and in the
hypothesis that the phase shift distribution is uniform
($\langle Tr S^n\rangle = 0$ for all $n \ne 0$) we have Eq. (\ref{denseigen2})
of the text

\begin{equation}
{\rm d}(E) =
\frac{M}{2\pi\hbar} \ \langle \tw(E)\rangle \ ,
\nonumber
\end{equation}

\nin where the Wigner time $\tw(E)$ is given by (\ref{tauw}) or

\begin{equation}
\tw(E) = \frac{1}{M} \sum_{m=1}^{M} \tau_{m}(E) \ .
\label{tauwapp}
\end{equation}

The Wigner time is the mean slope of the curves $\theta_{m}(E)$ in Fig. 10 and
has the physical interpretation of the mean time spent by the
particles in the scattering region, as can be easily seen from our
semiclassical analysis of the transmission amplitude (Appendix \ref{sec:Semi}).
Notice that the Wigner time resulting from our convention for
the $S$ matrix is not the usual delay time but and absolute
(always positive) time. In particular, for a perfect
(non-disordered) sample it is not zero, but given by $\tw =
2 \hbar/M \sum_{n=1}^{N} d(k_{n}L_{x})/dE =
\frac{1}{N} \sum_{n=1}^{N} L_{x}/v_{n}$, that is, the
average over the propagating modes of the ballistic traversal
times.

 From its definition (\ref{K2}) and the relationship (\ref{denseigen}) we can
write the density-density correlation function of the energy levels as

\begin{eqnarray}
K_{2}(E,E+\varepsilon)
= \frac{1}{(2\pi\hbar)^2} \sum_{m,m'=1}^{M} \sum_{n,n'=-\infty}^{\infty}
(-1)^{n+n'}
\nonumber \\
\langle \tau_{m}(E) \tau_{m'}(E)
\exp{(i(\theta_{m}(E)n-\theta_{m'}(E)n'-\tau_{m}(E)n'\varepsilon/\hbar)}
\rangle
- {\rm d}(E){\rm d}(E+\varepsilon) \ ,
\label{K2appen}
\end{eqnarray}

\nin where we have used $\theta_{m'}(E+\varepsilon) \approx
\theta_{m'}(E)+\tau_{m'}(E)\varepsilon/\hbar$ for $\varepsilon \ll E$
and we assumed that the $\tau_{m}(E)$ {\it are smooth functions
(in the scale of $\varepsilon$) which can be taken outside the
ensemble average}. If the phase shifts are uniformely distributed
only the terms with $n=n'$ survive the ensemble average and \cite{Bog,DoSm}

\begin{equation}
K_{2}(\varepsilon) = {\rm d}(E)\delta(\varepsilon)+{\rm d}^2(E) \left(
\frac{2 \pi}{M(M-1)} R_{2}\left(\frac{2 \pi}{M} \varepsilon {\rm d}(E) \right)
- 1 \right) \ .
\label{K2b}
\end{equation}

Scaling the energy difference with the mean level spacing
($e=\varepsilon
{\rm d}(E)$), in the large $M$ limit we obtain (\ref{K2c}) of the text
implying that {\it the phase shifts and the energy levels have the same
two-point correlation functions} when expressed in terms of their
respective reduced variables ($r$ and $e$). Notice that the
relationship between the phase shift correlations (given by $R_{2}(r)$
 or $Y_{2}(r)$) and the energy level correlations
(given by $K_{2}(e)$) is obtained in two steps:
first the quantization criterion
(\ref{quantcond}) (which neglects the evanescent modes) and second
assumptions on the homogeneity of the phase shift distribution and
the smooth character of $\tau_{m}(E)$.

\section {Semiclassical Approach}
\label{sec:Semi}

In this appendix we present a semiclassical treatment of the scattering in
a wave-guide geometry with disorder, for
understanding of the non-zero average of the diagonal reflection
amplitudes and of the various time scales in the problem.
The semiclassical
approximation to the transmission and reflection amplitudes can be obtained
by replacing the Green functions $G_{k}$ of Eqs. (\ref{alltrs}) by their
semiclassical path-integral expression \cite{gutz}

\begin{equation}
G^{\rm scl}_{k}(y^{\prime},y)=\frac{2\pi}{(2\pi i\hbar)^{3/2}}
\sum_{s(y,y^{\prime})} \sqrt
{D_{s}} \exp{\left(\frac{i}{\hbar}S_{s}(y^{\prime},y,E_f)-i
\frac{\pi}{2}\mu _{s}\right)} \ .
\label{eq:gfgutz}
\end{equation}

\nin The sum is over classical trajectories $s$ between the arguments of
the Green function (we will be omiting the $x$ dependence since it will be
assumed $x=0$ or $L_{x}$ depending whether we are considering reflection or
transmission),
$S_{s}$ is the action integral along the path $s$
at energy $E_f = \hbar^2 k^2/2 m = m v^2/2$, the stability prefactor is
$D_{s}=(v|\cos{\theta^{\prime}}|/m)^{-1}\mid \! (\partial
\theta /\partial y^{\prime})_{y} \! \mid$, $\theta$ and $\theta^{\prime}$ are
the
incoming and outgoing angles, and $\mu$ is the Maslov index given by the
number of constant-energy conjugate points \cite{gutz}. Performing the
transverse integrations by stationary-phase approximation, valid in the
$\hbar \to 0$ limit applicable to the many-channel case that we are
interested in, we obtain \cite{BJS}

\begin{equation}
r^{\rm scl}_{ba}=-\frac{\sqrt{2\pi i\hbar}}{2L_{y}}
\sum_{s(\bar{a},\bar{b})} {\rm sgn}(\bar{a}) \ {\rm sgn}(\bar{b}) \ \sqrt
{\tilde{D}_{s}} \
\exp{\left(\frac{i}{\hbar}\tilde{S}_{s}(\bar{b},\bar{a},E_{f})
-i \frac{\pi}{2}{\tilde\mu}_{s}\right)} \ ,
\label{eq:semiclass}
\end{equation}

\nin where the sum is taken over trajectories $s$ between the cross section
at $x=0$ with incoming and outgoing angles $\theta$ and $\theta^{\prime}$
such that $\sin{\theta}=\bar{a}\pi/kL_{y}$ and
$\sin{\theta^{\prime}}=\bar{b}\pi/kL_{y}$ ($\bar{a}=\pm a$,
$\bar{b}=\pm b$). Since $\bar{a}$ and $\bar{b}$ are integers we see that the
semiclassical approximation yields the intuitive result that only
trajectories which enter and exit at discrete angles corresponding to the
allowed quantized transverse momenta contribute to reflection. The reduced
action (virial) is

\begin{equation}
\tilde{S}(\bar{b},\bar{a},E_f)=S(y_{0}^{\prime},y_{0},E_f)+\hbar \pi \bar{a}
y_{0}/L_{y} -\hbar \pi \bar{b} y_{0}^{\prime}/L_{y}  \ ,
\label{eq:virial}
\end{equation}

\nin the new pre-exponential factor is

\begin{equation}
\tilde{D}_{s}=\frac{1}{mv|\cos{\theta^{\prime}}|} \left|
\left(\frac{\partial y}{\partial \theta^{\prime}}\right)_{\theta}\right| \ ,
\label{eq:dtilt}
\end{equation}

\nin and the Maslov index $\tilde{\mu}$ is given by $\mu$ and the
signs of the second derivatives of the action (see Ref. \cite{BJS}). A similar
expression holds for the transmission amplitudes, but the trajectories are
now taken from the cross-section at $x=0$ to that at $x=L_{x}$. The Kronecker
$\delta_{a,b}$ of the reflection amplitude, Eq. (\ref{trb}), has been taken
care off by trajectories going from
$y$ to $y'$ along the cross section $x=0$ (without entering the disordered
region) not included in Eq. (\ref{eq:semiclass}) \cite{BJS}. The rapid increase
of the mean value of the reflection amplitude for high modes can be traced to
the $\cos{\theta'}$ of Eq. (\ref{eq:dtilt}) since $\theta'$ approaches $\pi/2$
as $a$ increases towards $N$.

When taking the average of the matrix elements of $S$
over different impurity configurations,  the long
trajectories will have a rapidly varying phase which
leads to zero average,
while the short reflection trajectories which do not explore
the bulk of the
sample will survive the average. In particular, trajectories corresponding
to high mode numbers will have a large injection angle which makes
very
likely their reflection within a mean free path $l$ from the entrance. The
absence of selfaveraging to zero for the diagonal reflection
elements subsists in the
limit of $L_{x} \rightarrow \infty$ of a semi-infinite strip,
 since it is a
consequence of fast (short trajectory) processes. These short processes are
the only ones contributing to $\langle Tr S \rangle$ in the large $N$-limit
and this is the reason why in the numerics and in Eq. (\ref{eq:Trpert}) the
trace of the scattering matrix depends only on the properties at the entrance
of the sample. However, the reflected trajectories have a wide distribution of
lengths, and we do not have as in Nuclear Physics \cite{Brink} a clear-cut
separation between direct (or fast) and long processes.

In order to put these ideas in a more quantitative level, we
calculate the semiclassical approximation to $Tr S^{n}$. For $n=1$ we only need
the trace of the reflection matrices $r$ and $r'$, which can be evaluated
with the aid of Poisson summation formula,

\begin{eqnarray}
{\rm Tr} \ r^{\rm scl} = \sum_{a=1}^{N} r_{aa} =
-\frac{\sqrt{2\pi i\hbar}}{2L_{y}}
\sum_{m=-\infty}^{\infty} \int_{1/2}^{N+1/2} \ du \ \exp{(2 \pi i m u)}
\sum_{s(\bar{u},\bar{\bar{u}})} {\rm sgn}(\bar{u}) \ {\rm sgn}(\bar{\bar{u}})
\nonumber \\
\sqrt {\tilde{D}_{s}}
\exp{\left(\frac{i}{\hbar}\tilde{S}_{s}(\bar{\bar{u}},
\bar{u},E_{f}) -i \frac{\pi}{2}{\tilde\mu}_{s}\right)} \ ,
\label{eq:trr}
\end{eqnarray}

\nin where $\bar{u},\bar{\bar{u}}=\pm u$, not necessarily an integer. The
stationary-phase condition for the integral over $u$ is

\begin{equation}
{\rm sgn}(\bar{\bar{u}}) \ y'_{0} - {\rm sgn}(\bar{u}) \ y_{0} = 2 m L_{y} \ .
\label{eq:statph}
\end{equation}

\nin Since $0 < y_{0}, y'_{0} < L_{y}$ the
only solution is $\bar{\bar{u}} = \bar{u}$, $y'_{0} = y_{0}$, and
$m=0$, i.e., the trajectory closes itself at the entrance cross-section
with a specular reflection, where the transverse momentum (in the $y$
direction)
is conserved. Keeping only the $m=0$ term corresponds to the approximation of
replacing the sum over $a$ in ${\rm Tr} \ r^{\rm scl}$ by an integral; the
stationary-phase argument indicates that this is indeed the leading term in
$\hbar$. We then have

\begin{equation}
{\rm Tr} \ r^{\rm scl} = -\frac{i \pi \hbar}{\sqrt{2}L_{y}}
\sum_{s(y_{0},\bar{u}_{0};y_{0},\bar{u}_{0})} \sqrt {\hat{D}_{s}} \
\exp{\left(\frac{i}{\hbar}S_{s}(y_{0},y_{0},E_{f}) -i \frac{\pi}{2}
{\hat \mu _{s}}\right)} \ ,
\label{eq:trrf}
\end{equation}

\nin $S_{s}$ is the action integral along the classical path $s$, the
stability prefactor is

\begin{equation}
{\hat{D}_{s}} = {\tilde{D}_{s}} \left| \left(\frac{\partial^2 {\tilde{S}_{s}}}
{\partial a^{2}}\right)_{b} + \left(\frac{\partial^2 {\tilde{S}_{s}}}
{\partial b^{2}}\right)_{a} \right| ^{-1}_{\bar{a}=\bar{b}=\bar{u}_{0}} \ ,
\label{eq:dhat}
\end{equation}

\nin and the new Maslov index ${\hat \mu_{s}}$ is given by ${\tilde \mu _{s}}$
and the sign of the second derivative of the reduced action. For the
semiclassical approximation to
${\rm Tr} \ r'$ we have a similar expression, given in terms of
trajectories who have a specular reflection at the $x=L_{x}$ cross-section.

Higher powers of $S$ involve products of the matrices $r,t,r',t'$. For
instance,
the reflection submatrix of $S^2$ is $rr + t't$, and represents the two
possible ways of returning to the entrance cross-section touching only once one
of the vertical boundaries. The matrix products of the reflection and
transmission amplitudes can also be obtained with the aid of Poisson summation
formula, resulting in a stationary-phase condition very similar to Eq.
(\ref{eq:statph}) (but now the terms with $y_{0}$ and $y'_{0}$ come from
actions
of different trajectories, i.e. one from $t$ and the other from $t'$) which
indicates that we also have specular
reflections at each bounce with the vertical walls. Once we take the trace we
arrive at the intuitive result that ${\rm Tr}S^n$ is given in a semiclassical
approximation by closed trajectories, of a system with hard walls at the
extreme cross-sections, with $n$ bounces with the vertical walls.

Bl\"umel and Smilansky \cite{BlSm} expressed ${\rm Tr}S^n$ semiclassically in
terms of periodic orbits of the Poincar\'e scattering map, while in Ref.
\cite{DoSm} it was shown that a periodic orbit of order $n$ of the mapping
corresponds to a periodic orbit of the closed system which hits $n$ times the
extreme cross-sections. This is precisely our statement of last paragraph.

The semiclassical approximation to ${\rm Tr}S^n$, given by closed
orbits hitting $n$ times the vertical hard walls, is very useful
since we can
identify high values of $n$ with long trajectories. The lower harmonics of
$\rho(\theta)$ are given by the ensemble average over short orbits, which is
not zero in general; while the higher harmonics are given by long
orbits leading to
self averaging. This explains why $\rho(\theta)$ is relatively smooth and
justifies the usefulness of our diagrammatic approach yielding only
$\rho_{1}(\theta)$.

According to Eq. (\ref{sn}) the two-level form factor is given by
the ensemble average of $|TrS^n|^2$, therefore the semiclassical approach
also allows us to understand the phase shift correlations.
The difference
in the semiclassical approaches to the mean density and the TLFF is that the
former is expressed as a one-phase sum while the latter a two-phase sum
given by products of two periodic orbits of order $n$. In a diagonal
approximation we would take both trajectories to be the same or symmetrically
related. This yields a factor of two difference between the cases with and
without magnetic field, which we obtain for small values of $n$ (or in the
slope at the origen) in our numerical simulations (see Fig. 2). Since the two
curves for the TLFF must have the same area, they have to
cross at one point, indicating that the diagonal approximation ceases to be
valid for long trajectories. Therefore, for relatively low $n$ we can
approximate the Fourier components $\snM$ as simply given by periodic orbits
hitting $n$ times the vertical cross sections.

The semiclassical approach allows us to give an intuitive interpretation
of the Wigner time. According to Eqs.~(\ref{tauw}) and (\ref{eq:semiclass})
we have

\be
\tw(E) = \frac{\hbar}{iM} \sum_{i,j}^{M} S_{ji}^{*} \frac{dS_{ji}}{dE} =
\frac{\hbar}{iN} \sum_{a,b}^{M} \sum_{s(\bar{a},\bar{b})}
\sum_{s'(\bar{a},\bar{b})} \
\sqrt {\tilde{D}_{s}\tilde{D}_{s'}} \ T_{s'}
\exp{\left(\frac{i}{\hbar}(\tilde{S}_{s}-\tilde{S}_{s'})
-i \frac{\pi}{2}({\tilde\mu}_{s}-{\tilde\mu}_{s'})\right)} \ .
\label{tauwsemi}
\ee

\nin The sum is over all pairs of trajectories $s$ and $s'$ contributing
to $r$, $r'$, $t$ or $t'$. In performing the ensemble average the
short trajectories (where the diagonal approximation is valid) will
dominate and for $s=s'$ we have

\begin{equation}
\langle \tw(E) \rangle = \frac{\hbar}{M}  \sum_{ab}^{M}
\sum_{s(\bar{a},\bar{b})} {\tilde{D}_{s}} \ T_{s} \ ,
\label{tausemiav}
\end{equation}

\nin giving the standard interpretation of the Wigner time as the mean
scattering time (mean time of the classical scattering trajectories
averaged with their stability prefactor) or the inverse escape rate
\cite{DSF} from the disordered region.

The criterion developed Sec. \ref{sec:Devi} was to measure the departures
from the circular ensembles by the parameter
$\eta_{c} = \tw E_{c}/\hbar$, which is simply the ratio between the
scattering and the diffusion times across the sample. The agreement with
RMT obtained for large values of $\eta_{c}$ translates into a Wigner time
much larger than the diffusing time for traversing the structure
($L_{x}^{2}/D$). That is, the particle spends in the disordered region enough
time to sample it before leaving.

 Separating the reflection and transmission trajectories $s$ of
Eq.~(\ref{tausemiav}) we see that, at the semiclassical level, the Wigner
time can be decomposed into transmission and reflection components. At the
quantum mechanical level, this is not possible since the matrices
$r$, $r'$, $t$ and $t'$ are not unitary, and for a given impurity realization
we obtain an imaginary part when using these matrices instead of $S$ in
Eq.~(\protect\ref{tauw}). However, these imaginary parts vanish under
ensemble average giving well-defined transmission and reflection Wigner
times. These partial times can be related with the classical traversal
times of a diffusion through a disordered region connected with leads
(diffusion from an entrance to an exit \cite{LanBut,Fisher}). Calling
$\tau_{e}=l/v_{f}$ the elastic scattering time, the quasi-one-dimensional
case can be discretized as a diffusion in a one-dimensional lattice with
$m=L_{x}/l$ sites. In this model it has been show \cite{LanBut,Fisher}
that the mean traversal and reflection times are given by

\begin{mathletters}
\label{alltdees}
\begin{equation}
\tau_{\scriptscriptstyle T} = \tau_{e} \ m \ \left( \frac{1}{3} \ m +
\frac{2}{3} \right) \ ,
\label{tdeea}
\end{equation}
\begin{equation}
\tau_{\scriptscriptstyle R} = \tau_{e} \ \left( \frac{2}{3} \ m +
\frac{1}{3} \right) \ .
\label{tdeeb}
\end{equation}
\end{mathletters}

\nin Therefore, in the large $m$-limit, the mean traversal time is reduced
by a factor of three with respect to the diffusion time in an infinite chain,
$\tau_{\scriptscriptstyle T} \approx 1/3 \ \tau_{e} \ m^{2} = 1/3
\ (L_{x}^{2}/2D)$. The classical transmission and reflection probabilities
are

\begin{mathletters}
\label{alltcs}
\begin{equation}
T = \frac{1}{m+1} \ ,
\label{tca}
\end{equation}
\begin{equation}
R = \frac{m}{m+1} \ ,
\label{tcb}
\end{equation}
\end{mathletters}

\nin and the total dwell time is

\begin{equation}
\tau_{\scriptscriptstyle D} = \tau_{\scriptscriptstyle T} T +
\tau_{\scriptscriptstyle R} R = \tau_{e} \ m = L_{x}/v_{f} \ ,
\label{tdt}
\end{equation}

\nin consistent with our findings that the Wigner time is
approximately independent of disorder
and coincides with the ballistic traversal time. For quasi-one-dimensional
geometries the partial Wigner times obtained in the simulations after
impurity average are in relatively good agreement with the average
diffusion times $\tau_{\scriptscriptstyle T} T$ and
$\tau_{\scriptscriptstyle R} R$. From Eqs.~(\ref{alltdees}) and
(\ref{alltcs}) we can also calculate the mean lenght of trajectories
contributing to $\snM$, that is, those hitting $n$ times the vertical walls
at the entrance of the disordered region. For instance, for $n=2$ we have

\begin{equation}
\tau_{2} = \frac{2 \tau_{\scriptscriptstyle T} T^{2} +
2 \tau_{\scriptscriptstyle R} R^{2}}{T^{2} + R^{2}}
\label{last}
\end{equation}

\nin In the large $m$-limit, trajectories with $n$ bounces will have a
typical length proportional to $n$.

\section {Invariant Measure of S in the Polar Decomposition without
time-reversal symmetry}
\label{sec:CUE}

In this appendix we find the metric of the scattering matrix in the polar
decomposition for the case where there is no time-reversal symmetry. The
calculation follows the same lines of that in Sec. \ref{sec:RelSM}, the
difference coming from the fact that instead of the decomposition
(\ref{eq:Sdecomp}) we now have

\begin{equation}
S = X^{\dagger} \Sigma \ X = U^{\prime} \ \Gamma \ U = Y^{\prime} \ Y \ .
\label{eq:SdecompD}
\end{equation}

\nin The unitary matrix $X$ diagonalizes $S$, $U$ and $U'$ are independent
block-diagonal unitary matrices given by the polar decomposition
(\ref{eq:Spol}). $Y$ and $Y'$ are independent unitary matrices that allow
to define the metric of $S$. Diagonalizing $\Gamma$ as in Eq.~(\ref{eq:Gamma})
we have the same $Y$ as before (Eq.~(\ref{eq:Umat})) and

\begin{equation}
Y' = U' \ O^{\rm T} \ F = \left( \begin{array}{lr}
u^{(3)} {\cal P} & \hspace{0.5cm} iu^{(3)}{\cal Q} \\
u^{(4)}{\cal Q} 	& \hspace{0.5cm} -iu^{(4)}{\cal P}
\end{array} \right) \ .
\label{eq:UmatD}
\end{equation}

The infinitesimal variations of $Y$, $u^{(1)}$, and $u^{(2)}$ are given by

(\ref{eq:dU}) and (\ref{eq:dvl}) in terms of the antihermitian matrices
$\delta Y$, $\delta u^{(1)}$ and $\delta u^{(2)}$. Analogously, we can
write

\begin{equation}
dY = Y' \ \delta Y' \ ,
\label{eq:dUD}
\end{equation}

\begin{equation}
du^{(l)} = u^{(l)} \ \delta u^{(l)} \ , \hspace{2cm} \delta u^{(l)} =
da^{(l)}+i \ ds^{(l)} \ ,
\hspace{2cm} l=3,4 .
\label{eq:dvlD}
\end{equation}

\nin Therefore the infinitesimal variations of $S$ are given by

\begin{equation}
dS = Y' \ (\delta Y' + \delta Y) \ Y = Y' \ (id\HT ) \ Y \ .
\label{eq:dSD}
\end{equation}

\nin The hermitian matrix $d\HT$ defines the metric of $S$ by \cite{Dy}

\begin{equation}
\mu(dS) = \prod_{i \leq j}^{2N} Re(d\HT)_{ij} \ \prod_{i < j}^{2N}
Im(d\HT)_{ij} \ ,
\label{eq:mudSD}
\end{equation}

\nin and is given by

\begin{eqnarray}
\hspace{-5cm} d\HT = 2 \ \left( \begin{array}{cc}
0 & \hspace{0.5cm} {\cal P} \ d{\cal Q}-{\cal Q} \ d{\cal P} \\
{\cal P} \ d{\cal Q}-{\cal Q} \ d{\cal P} & 0
\end{array} \right) \ +
\nonumber \\
\nonumber \\
+ \left( \begin{array}{lr}
{\cal P} (d s^{(1)}+d s^{(3)}) {\cal P}+
{\cal Q} (d s^{(2)}+d s^{(4)}) {\cal Q} &
\hspace{0.5cm}
-{\cal P} (d a^{(1)} - d a^{(3)}) {\cal Q}+
{\cal Q} (d a^{(2)} - d a^{(4)}) {\cal P} \\
{\cal Q} (d a^{(1)} - d a^{(3)}) {\cal P}-
{\cal P} (d a^{(2)} - d a^{(4)}) {\cal Q}
& \hspace{0.5cm}
{\cal Q} (d s^{(1)} + d s^{(3)}) {\cal Q}+
{\cal P} (d s^{(2)} + d s^{(4)}) {\cal P}
\end{array} \right) \ -
\nonumber \\
\nonumber \\
- i \ \left( \begin{array}{lr}
{\cal P} (d a^{(1)}+d a^{(3)}) {\cal P}+
{\cal Q} (d a^{(2)}+d a^{(4)}) {\cal Q} &
\hspace{0.5cm}
{\cal P} (d s^{(1)} - d s^{(3)}) {\cal Q}-
{\cal Q} (d s^{(2)} - d s^{(4)}) {\cal P} \\
-{\cal Q} (d s^{(1)} - d s^{(3)}) {\cal P}+
{\cal P} (d s^{(2)} - d s^{(4)}) {\cal Q}
& \hspace{0.5cm}
{\cal Q} (d a^{(1)} + d a^{(3)}) {\cal Q}+
{\cal P} (d a^{(2)} + d a^{(4)}) {\cal P}
\end{array} \right) \ .
\label{eq:dAmatD}
\end{eqnarray}

In order to calculate the invariant measure of $S$ it is convenient to break
the product (\ref{eq:mudSD}) over the real and imaginary parts of the elements
of $d\HT$ as

\begin{mathletters}
\label{allbtds}
\begin{equation}
\prod_{a=1}^{N} Re(d\HT)_{a,a+\Ns} = \prod_{a=1}^{N}
\frac{1}{2 \sqrt{\lambda_{a}} (1+\lambda_{a})} \
d \lambda_{a} \ ,
\label{btda}
\end{equation}
\begin{equation}
\prod_{a < b}^{N} Im(d\HT)_{a,b} \ Im(d\HT)_{a+\Ns,b+\Ns} \
Re(d\HT)_{a,b+\Ns} \ Re(d\HT)_{b,a+\Ns} =
\prod_{a < b}^{N} \left(\frac{\lambda_{a}}{\lambda_{a}+1} -
\frac{\lambda_{b}}{\lambda_{b}+1} \right) \prod_{l=1}^{4} da^{(l)}_{ab} \ ,
\label{btdb}
\end{equation}
\begin{equation}
\prod_{a < b}^{N} Re(d\HT)_{a,b} \ Re(d\HT)_{a+\Ns,b+\Ns} \
Im(d\HT)_{a,b+\Ns} \ Im(d\HT)_{b,a+\Ns} =
\prod_{a < b}^{N} \left(\frac{\lambda_{a}}{\lambda_{a}+1} -
\frac{\lambda_{b}}{\lambda_{b}+1} \right) \prod_{l=1}^{4} ds^{(l)}_{ab} \ ,
\label{btdc}
\end{equation}
\begin{eqnarray}
\prod_{a=1}^{N} Re(d\HT)_{a,a} \ Re(d\HT)_{a+\Ns,a+\Ns} \ Im(d\HT)_{a,a+\Ns} =
\prod_{a=1}^{N}
\left({\cal P}^{2}_{a} \ (d s^{(1)}_{aa}+d s^{(3)}_{aa}) +
{\cal Q}^{2}_{a} \ (d s^{(2)}_{aa}+d s^{(4)}_{aa}) \right) \times
\nonumber \\
\left({\cal Q}^{2}_{a} \ (d s^{(1)}_{aa}+d s^{(3)}_{aa}) +
{\cal P}^{2}_{a} \ (d s^{(2)}_{aa}+d s^{(4)}_{aa}) \right) \ \times
\left({\cal P}_{a}{\cal Q}_{a} \ (d s^{(1)}_{aa}-d s^{(3)}_{aa}) -
{\cal Q}_{a}{\cal P}_{a} \ (d s^{(2)}_{aa}-d s^{(4)}_{aa}) \right) \ .
\label{btdd}
\end{eqnarray}
\end{mathletters}

The last product, Eq.~(\ref{btdd}), shows the redundancy of the polar
decomposition in the absence of time-reversal symmetry. The matrix $S$
is unchanged if in its polar decomposition we make the transformation
\cite{MS}

\begin{equation}
U \rightarrow GU \ , \hspace{2cm} U' \rightarrow U'G^{\dagger} \ ,
\hspace{2cm} G = \left( \begin{array}{lr}
{\cal G}	& \hspace{0.5cm} 0	\\
0	& \hspace{0.5cm} {\cal G}
\end{array} \right) \ ,
\label{eq:trapo}
\end{equation}

\nin where the blocks ${\cal G}$ are pure-phase diagonal matrices,
${\cal G}_{a}=\exp{(i\eta_{a})}$. Under this transformation we have

\begin{equation}
ds^{(l)}_{aa}  \rightarrow ds^{(l)}_{aa} \pm d \eta_{a} \ , \hspace{2cm}
l=1,2 \ (3,4) \,
\label{trapos}
\end{equation}

\nin and the r.h.s. of Eq.~(\ref{btdd}) remains unchanged. Since
$d \eta_{a}$ are arbitrary we take them equal to $ds^{(1)}_{aa}$ \cite{MS},
and when multiplying both sides of the equation above by $\prod_{a=1}^{N}
ds^{(1)}_{aa}$ we obtain the transformation of the volume element as

\begin{equation}
\prod_{a=1}^{N} Re(d\HT)_{a,a} \ Re(d\HT)_{a+\Ns,a+\Ns} \ Im(d\HT)_{a,a+\Ns} \
ds^{(1)}_{aa} = \prod_{a=1}^{N} \left(
\frac{\sqrt{\lambda_{a}}}{\lambda_{a}+1} \right)
\prod_{l=1}^{4} ds^{(l)}_{aa} \ .
\label{volvol}
\end{equation}

As in the time-symmetric case, we express the jacobian of the transformation
in the Gibbs form (\ref{eq:jacobian}), with an inverse effective temperature
$\beta=2$ and an effective hamiltonian (\ref{eq:hamiltonian}) given by the
same logarithmic interaction as before (Eq.~(\ref{eq:interaction})) and
a one-body potential:

\begin{equation}
V(\lambda) = N \ \log{(1+\lambda)} \ ,
\label{eq:potentialD}
\end{equation}

\nin which coincides with (\ref{eq:potential}) up to order $N$ terms but
differs in the terms of order unity ($N^{0}$).

\newpage

\begin{table}
\begin{tabular}{|cccccccccccccccc|}
\ run \		& \ $\phi/\phi_{0}$ \     & \ $N\!S$ \	& \ $AR$ \
	& \ $L_{y}$ \  & \ $N$ \  & \ $E_{f}$ \		& \ $k$ \
	& \ $W$ \	&$\frac{\tau_{W}\Delta}{\hbar}\frac{N}{\pi}$
	& \ $l$	\ 	& \ $r_c$ \	& \ $\phi_{c}$ \
	& \ $\langle g \rangle \ $	& \ $\langle \delta g^2 \rangle$ \
	& \ $|\langle Tr S \rangle|$ \ \\ \hline\hline
R1	&0	&5000	&4	&34   &14  &-2.5	&1.32	&1	&1.20
	&28	&0.67	&0.007	&4.14	&0.13   &0.25 \\
F1	&2	&5000	&	&     &    &	        &	&	&
	&	&	&	&4.23	&0.09   & \\ \hline
R2	&0	&1400	&4	&34   &14  &-2.5	&1.32	&1.5	&1.24
	&13	&0.32	&0.028	&2.13	&0.13   &0.38 \\
F2	&2	&1400	&	&   &  &	&	&	&	&
	&	&	&2.23	&0.08   & \\ \hline
R3	&0	&700	&4	&34   &14  &-2.5	&1.32	&2	&1.20
	&8.3	&0.20	&0.042	&1.11	&0.11       &0.51  \\
F3	&2	&700	&	&   &  &	&	&	&	&
	&	&	&1.26		&0.08   & \\ \hline
R4	&0	&2500	&4	&34   &14  &-2.5	&1.32	&4	&1.24
	&2.8	&	&	&0.02	&4.98   &1.68 \\
F4	&2	&2500	&	&   &  &	&	&	&	&
	&	&	&0.03	&4.25   & \\ \hline
R5	&0	&700	&4	&34   &14  &-2.5	&1.32	&6	&0.34
	&1.8	&	&	&$10^{-7}$	&24.9   &3.60 \\
F5	&2	&700	&	&   &  &	&	&	&	&
	&	&	&$2 \! \times \! 10^{-7}$	&24.8   & \\ \hline
R6	&0	&700	&4	&34  &20  &-1.36	        &1.90	&1.5
	&1.60	&12	&0.37	&0.028	&2.62	&0.13   &0.15 \\ \hline
R7	&0	&500	&7	&34  &20  &-1.36	        &1.90	&1
	&1.53	&27	&0.37	&0.025	&3.47	&0.13   &0.13 \\ \hline
R8	&0	&1400	&10	&34  &14  &-2.5	        &1.32	&1	&1.21
	&28	&0.26	&0.033	&1.91	&0.13   &0.20 \\
F8	&5	&1400	&	&     &   &             &	&	&
	&	&	&	&2.07	&0.07   & \\ \hline
R9	&0	&700	&10	&34  &20  &-1.36	        &1.90	&1
	&1.45	&27	&0.30	&0.028	&2.50	&0.13   &0.19 \\ \hline
R10	&0	&700	&30	&34  &14  &-2.5	        &1.32	&1	&1.27
	&28	&	&	&0.45	&0.58   &0.28 \\
F10	&15	&700	&	&     &    &	&	&	&       &
	&	&	&0.69   &0.22 	& \\ \hline
R11	&0	&600	&4	&68   &14  &-3.56	&0.68	&1	&1.11
	&24	&0.34	&0.036	&1.85	&0.13   &1.14 \\  \hline
R12	&0 	&2500	&1	&68   &14  &-3.56	&0.68	&1	&1.12
	&24	&1.35	&0.012	&5.49	&0.18   &1.15 \\
F12	&2	&1400	&	&   &  &	&	&	&	&
	&	&	&5.55	&0.15   & \\ \hline
R13	&0	&700	&1	&68   &14  &-3.56	&0.68	&1.5	&1.10
	&11	&0.60	&0.045	&3.19	&0.18   &2.16 \\ \hline
R14	&0	&700	&1	&68   &14  &-2.5	&0.68	&2	&0.97
	&5.9	&0.30	&0.093	&1.96	&0.17   &3.28 \\ \hline
R15	&0	&700	&1	&68   &20  &-3.17	&0.95	&1.5	&1.14
	&13	&0.98	&0.021	&5.22	&0.18   &1.48 \\ \hline
R16	&0	&700	&1	&68   &20  &-3.17	&0.95	&2.0	&1.16
	&7.1	&0.54	&0.060	&3.27	&0.17   &2.37 \\ \hline
R17	&0	&500	&1	&68   &25  &-2.76	&1.18	&1.5	&1.21
	&14	&1.30	&0.015	&6.86	&0.19   &1.00 \\ \hline
R18	&0	&400	&1	&68   &28  &-2.5	&1.32	&1.5	&1.25
	&14	&1.41	&0.016	&7.68	&0.19   &0.93 \\ \hline
R19	&0	&600	&1/4	&140   &14  &-3.90	&0.33	&1	&1.05
	&11	&0.08	&0.068	&6.12	&0.27   &4.36 \\
F19	&2	&200	&	&     &    &	        &	&	&
	&	&	&	&6.13	&0.09   & \\
\end{tabular}

\vspace{4ex}

\caption{
Runs with zero magnetic field are indicated by R, runs with a nonzero flux
$\phi$ through the sample are indicated by F ($\phi_{0}=hc/e$ is the elemental
flux). $N\!S$ is the number of samples (impurity configurations) considered.
$AR=L_{x}/L_{y}$ is the aspect ratio of the samples and $N$ is the number of
propagating modes. $E_{f}$ and $k$ are the Fermi energy and the
wavevector in Anderson units ($E_{f}=-2-2\cos{k}$). $W$ is the amplitude
of the on-site disorder of the run. $\protect\tw$ is the Wigner time and
$\Delta$ is the level spacing of a perfect (non-disordered) rectangle with
sides $L_{x}$, $L_{y}$. The column showing
$\frac{\tau_{W}\Delta}{\hbar}\frac{N}{\pi}$ checks the approximate validity
of Eq.~(\protect\ref{denseigen2}). $l$ is the elastic-mean-free path and
$r_c$ is the critical angle $\eta_{c}$ (Eq.~(\protect\ref{etac}))
scaled with the mean phase shift spacing. $\phi_{c}$ is the
slope of $\sigma^{2}$ evaluated at $r_{c}$ obtained from Fig. 8.
$\langle g \rangle$ is the averge dimensionless conductance and
$\langle \delta g^2 \rangle$ its variance. $|\langle Tr S \rangle|$ is the
average trace of the scattering matrix.
}
\end{table}

\newpage

\begin{figure}
\caption{
Typical sample geometry where the disordered scattering
is confined to the
region $0 \leq x \leq L_{x}$ of the wave guide. $A$ and $D$ are the amplitudes
of the incoming flux, $B$ and $C$ are the outgoing amplitudes.
}
\end{figure}

\begin{figure}
\caption{
Phase shift form factor (Eq.\ (\protect\ref{sn})) for
quasi--$1d$ conductors (upper inset) without magnetic field (R1)
and with $\phi/\phi_{0}=2$ (F1).
The large-$M$ OE and UE values are given by thick solid and dashed lines. The
number of impurity configurations is $N\!S=5000$, the aspect ratio is $AR=4$,
the disorder is $W=1$ and the number of propagating modes is $N=M/2=14$. Lower
inset: histograms of the phase shift distribution normalized by the uniform
value $\rho_{0}=M/2 \pi$. The
R1 distribution is off-set by $-1/3$, while the F1 distribution
is off-set by $-2/3$.
}
\end{figure}

\begin{figure}
\caption{
Phase shift form factor for quasi--$1d$ insulators without (a) and
with (b) magnetic field. Filled circles correspond to strong
localization, while diamonds represent a quasi--$1d$ metal--insulator
crossover ($L_x \approx \xi$) with weak disorder.
Histograms in the insets show the
corresponding phase shift distributions with an off-set of -1/3 for
the strong disorder case and an
off-set of -2/3 for weak disorder. The thick solid and dash lines
in a (b) represent respectively the large-$M$ form factor of one OE
(UE) and of the superposition of two independent OE (UE).
}
\end{figure}

\begin{figure}
\caption{
Number variance, with (a) and without (b) time-reversal symmetry,
as we approach the localized regime by increasing the disorder or
the system length. Thick solid lines correspond to the number
variance in the
orthogonal (unitary) ensembles,
Eq.\ (\protect\ref{alles}), while thick dashed lines correspond to the
number variance of the superposition ensembles,
Eq.\ (\protect\ref{Sigmas}). White symbols show the approach to the
quasi--$1d$ localized regime by increasing the length keeping the
disorder weak
($W=1$): $AR=4$ (squares), 10 (circles) and 30 (diamonds). Filled
symbols
show the approach to the localized regime by increasing the disorder
keeping the geometry fixed ($AR=4$), $W=2$ (triangles), 4 (circles), 6
(squares).
}
\end{figure}

\begin{figure}
\caption{
Diagrams of second-order perturbation theory for the average
Green function used to calculate the average of the reflection matrix
elements.
}
\end{figure}

\begin{figure}
\caption{
Average values of the diagonal reflection amplitudes as a function of the mode
number for the samples scketched in the insets. The solid thick lines are
guide-to-the eye with the functional form $1/k_{a}$ and $N$=14 modes.
(a) R1: $AR=4$, $W$=1; R8: $AR=10$, $W$=1; R2: $AR=4$, $W=1.5$.
(b) R12: $AR=1$, $W$=1, $N$=14; R11: $AR=4$, $W$=1; R13: $AR=1$, $W=1.5$,
$N=14$; R18: $AR=1$, $W$=1.5, $N$=28. For the last sample we show the
average diagonal reflection amplitude for only the first half of the mode
numbers.
}
\end{figure}

\begin{figure}
\caption{
(a) Phase shift form factors after unfolding the raw data
for square samples (upper inset) for $N=14$ propagating modes.
The sample with
a disorder strenght of $W=1$ is represented by squares (R12) for $B=0$ and
triangles (F12) for $\phi/\phi_{0}=2$. Circles represent a sample with
$W=1.5$ and no magnetic field (R13). The large-$M$ OE and UE values are
given by solid and dashed lines. Lower inset: histograms of the
phase shift distribution normalized by the uniform value $\rho_{0}=M/2 \pi$.
The
distribution R12 (F12) is off-set by -0.5 (-0.75), while the distribution of
R13 is off-set by -0.25.
(b) Phase shift form factors after unfolding the raw data
(circles) and according to Eq.\ (\protect\ref{sn}) (filled squares) for
slab-shape samples (upper inset) without magnetic field (R19). Triangles:
Form factors according to Eq.\ (\protect\ref{sn}) for $\phi/\phi_{0}=2$
(F19). The large-$M$ OE and UE values are given by solid and dashed lines.
Lower inset: histograms of the phase shift
distribution normalized by the uniform value $\rho_{0}=M/2 \pi$. The
R19 distribution is off-set by -0.4, while the F19 distribution is off-set
by -0.8.
}
\end{figure}

\begin{figure}
\caption{
Difference between the number variance obtained in the numerical
simulations and the number variance of the standard ensembles.
$\sigma^{2}(r) =
\Sigma^{2}(r) - \Sigma^{2}_{OE}(r)$ for samples without magnetic field and
$\sigma^{2}(r) = \Sigma^{2}(r) - \Sigma^{2}_{UE}(r)$ for samples with
magnetic field. Solid (dashed) lines are for samples without (with)
magnetic field. Thin lines are for square geometries, R12, F12: $AR=1$, $W$=1,
$N=14$; R13: $AR=1$, $W=1.5$, $N=14$; R15: $AR=1$, $W=1.5$, $N=20$. Thick
lines are for thin slabs, R19, F19: $AR=1/4$, $W$=1, $N$=14.
}
\end{figure}

\begin{figure}
\caption{
Slope of $\sigma^{2}(r)$ (obtained from Figs. 6 and 8) as a function of
the inverse of the reduced critical angle
$r_c = (\tw E_{c}/\hbar) \ (M/2\pi)$. The symbols used for the various runs
are consistent with the labels of Table I of Appendix A. Samples
R1-R11 are quasi-one dimensional, while samples R12-R17 represent
the two-dimensional case (square geometries). The dashed
straight lines are guide-to-the-eye through the two groups of
data points and illustrate the approximate validity of
Eq.~(\protect\ref{eq:slope}).
}
\end{figure}

\begin{figure}
\caption{
Energy dependence of the phase shifts for one of the quasi--$1d$
conductors
studied in Fig. 1 (aspect ratio $AR=4$, disorder
$W=1$) in an energy range where there are $N=14$ propagating modes.
}
\end{figure}


\begin{references}
\bibitem{PMPC} This remarkable fact, which can be shown by a direct
evaluation of the conductance fluctuations from the statistics of matrix
elements $S_{ij}$
was brought to our attention by P.~Mello (Private communication).
\end{references}
\end{document}